\documentclass[%
 aip,
 amsmath,amssymb,
 reprint,%
]{revtex4-1}

\usepackage{graphicx}% Include figure files
\usepackage{dcolumn}% Align table columns on decimal point
\usepackage{bm}% bold math
\usepackage[utf8]{inputenc}
\usepackage[T1]{fontenc}
\usepackage{mathptmx}
\usepackage{etoolbox}
\usepackage{booktabs,caption}

\makeatletter
\def\@email#1#2{%
 \endgroup
 \patchcmd{\titleblock@produce}
  {\frontmatter@RRAPformat}
  {\frontmatter@RRAPformat{\produce@RRAP{*#1\href{mailto:#2}{#2}}}\frontmatter@RRAPformat}
  {}{}
}%
\makeatother
\begin{document}

\preprint{AIP/123-QED}

\title{Phase-selective growth of $\kappa$- vs $\beta$-Ga$_2$O$_3$ and (In$_x$Ga$_{1-x}$)$_2$O$_3$ by In-mediated metal exchange catalysis in plasma-assisted molecular beam epitaxy}
% Force line breaks with \\
\author{A. Ardenghi*}
 
\author{O. Bierwagen*}%
 \email{bierwagen@pdi-berlin.de}
 \author{J. L{\"a}hnemann}
\affiliation{Paul-Drude-Institut für Festkörperelektronik, Leibniz-Institut im Forschungsverbund Berlin e.V., Hausvogteiplatz 5-7, 10117 Berlin, Germany}%

\author{J. Kler}
\author{A. Falkenstein}
\author{M. Martin}
\affiliation{Institute of Physical Chemistry, RWTH Aachen University, D-52056 Aachen, Germany}%

\author{P. Mazzolini*}
\email{piero.mazzolini@unipr.it}
\affiliation{Department of Mathematical, Physical and Computer Sciences, University of Parma, Parco Area delle Scienze 7/A, 43124 Parma, Italy}%

\email{ardenghi@pdi-berlin.de}

\begin{abstract}

Its piezo- and potentially ferroelectric properties make the metastable kappa polymorph of Ga$_2$O$_3$ an interesting material for multiple applications, while In-incorporation into any polymorphs of Ga$_2$O$_3$ allows to lower their bandgap. In this work, we provide a guideline to achieve single phase $\kappa$-, $\beta$-Ga$_2$O$_3$ as well as their (In$_x$Ga$_{1-x}$)$_2$O$_3$ alloys up to $x$ = 0.14 and $x$ = 0.17 respectively, using In-mediated metal exchange catalysis in plasma assisted molecular beam epitaxy (MEXCAT-MBE). The polymorph transition from $\kappa$ to $\beta$ is also addressed, highlighting the fundamental role played by the thermal stability of the $\kappa$-Ga$_2$O$_3$. Additionally, we also demonstrate the possibility to grow ($\bar{2}$01) $\beta$-Ga$_2$O$_3$ on top of $\alpha$-Al$_2$O$_3$ (0001) at temperatures at least 100 °C above those achievable with conventional non-catalyzed MBE, opening the road for increased crystal quality in heteroepitaxy. The role of the substrate, as well as strain and structural defects in the growth of $\kappa$-Ga$_2$O$_3$ is also investigated by growing simultaneously on three different materials: \textit{(i)} $\alpha$-Al$_2$O$_3$ (0001), \textit{(ii)} 20 nm of ($\bar{2}$01) $\beta$-Ga$_2$O$_3$ on $\alpha$-Al$_2$O$_3$ (0001) and \textit{(iii)} ($\bar{2}$01) $\beta$-Ga$_2$O$_3$ single crystal.
\end{abstract}
\maketitle

\section{\label{sec:Int}Introduction}

Lately, metal oxides have gained a lot of attention due to their potential in electronic applications. One of the most researched materials is monoclinic gallium oxide ($\beta$-Ga$_2$O$_3$), an ultra-wide bandgap semiconductor (\textit{E$_g$} $\approx$ 4.8 eV) which may have potential as the future of high voltage power devices.\cite{doi:10.1063/5.0060327} $\beta$-Ga$_2$O$_3$ can be grown from the melt,\cite{galazkaScalingUpBulkVGa2017,kuramataHighqualityVGaSingle2016a,ohbaGrowthVGa2O3Single2021a,villoraLargesizeVGa2O3Single2004} its electrical properties can be modulated through n-type doping\cite{doi:10.1063/5.0027870,Ahmadi_2017,doi:10.1063/5.0087987,doi:10.1063/5.0101132, SiO2doping} and its bandgap can be engineered through Al\cite{doi:10.1063/5.0131285} and In\cite{doi:10.1021/acsami.0c16021} alloying. Gallium oxide can also be grown in other crystal structures. Besides the thermodynamically stable $\beta$-Ga$_2$O$_3$, metastable phases, such as: orthorhombic ($\kappa$), corundum ($\alpha$), defect spinel ($\gamma$), and cubic ($\delta$) can be synthesized.

Sharing some material properties with the $\beta$-phase (e.g., similar \textit{E$_g$}), the $\kappa$-phase also represents another interesting candidate for new generation electronic devices. One of the main points of interest for $\kappa$-Ga$_2$O$_3$ is its suggested ferroelectricity\cite{mezzadriCrystalStructureFerroelectric2016} and high spontaneous polarization along the [001] direction,\cite{maccioniPhaseDiagramPolarization2016a} potentially leading to the creation of a controllable high density two-dimensional electron gas in properly designed heterostructure interfaces.\cite{rangaHighlyTunablePolarizationengineered2020, choEpitaxialEngineeringPolar2018,POLYAKOV2023168315, kuangBandAlignmentEnhanced2021} 
The biggest challenges of the $\kappa$-Ga$_2$O$_3$ polymorph are (i) its stabilization and (ii) the fine control of its electronic transport. As for (ii), several efforts are currently directed in the suppression/control of structural defects (e.g., rotational domains and extended defects\cite{mazzoliniSilaneMediatedExpansion2023, oshimaEpitaxialGrowthPhasepure2015, doi:10.1021/acsomega.0c04634}) which are the main limit for the electrical transport. For (i), stabilizing the $\kappa$-phase is challenging considering the lack of bulk substrates, making heteroepitaxy via both physical- or chemical-vapor deposition techniques (PVD, CVD) the only viable option.\cite{D0TC02743J}
Different substrates have been employed to stabilize (001)-oriented $\kappa$-Ga$_2$O$_3$, mostly GaN (0001),\cite{oshimaEpitaxialGrowthPhasepure2015} AlN (0001),\cite{oshimaEpitaxialGrowthPhasepure2015} MgO (111),\cite{kneissTinassistedHeteroepitaxialPLDgrowth2019,nishinakaHeteroepitaxialGrowthEGa2016} YSZ (111)\cite{kneissTinassistedHeteroepitaxialPLDgrowth2019,nishinakaHeteroepitaxialGrowthEGa2016} and Al$_{2}$O$_{3}$ (0001)\cite{vogtMetalExchangeCatalysisGrowth2017,krachtTinAssistedSynthesisGa2017,kneissTinassistedHeteroepitaxialPLDgrowth2019,boschi_hetero-epitaxy_2016,mezzadriCrystalStructureFerroelectric2016}. 

Differently from CVD, the synthesis of $\kappa$-Ga$_2$O$_3$ through PVD techniques (e.g., molecular beam epitaxy MBE and pulsed laser deposition PLD) requires the introduction of a metal catalyst (\textit{i.e.}, In\cite{vogtMetalExchangeCatalysisGrowth2017} or Sn\cite{krachtTinAssistedSynthesisGa2017,kneissTinassistedHeteroepitaxialPLDgrowth2019}) resulting in a peculiar growth mechanism defined as metal exchange catalysis (MEXCAT)\cite{mazzoliniSubstrateorientationDependenceVGa2020,mazzolini_offcut-related_2020} or metal-oxide catalyzed epitaxy (MOCATAXY).\cite{doi:10.1063/5.0101132,kargEnhancedEpitaxialGrowth2022a,mccandlessGrowthAGaAAl2023,mauzeMetalOxideCatalyzed2020,IndiumSurfKarg} It has also been recently demonstrated\cite{kargEnhancedEpitaxialGrowth2022a} that even the use of an ultrathin layer of SnO$_2$ leads to the stabilization of the $\kappa$-phase, without the need of a constant catalyst supply during growth. The importance of the deposition conditions in the stabilization of the phase has already been highlighted both with Sn-assisted MEXCAT growth\cite{kneissTinassistedHeteroepitaxialPLDgrowth2019,krachtTinAssistedSynthesisGa2017} of $\kappa$-Ga$_2$O$_3$ as well as $\kappa$-(Al$_x$Ga$_{1-x}$)$_2$O$_3$.\cite{hassaControlPhaseFormation2020} In addition, MEXCAT has been shown to also allow for increased growth rates of $\alpha$-Ga$_2$O$_3$ on m-plane Al$_2$O$_3$ substrates\cite{mccandlessGrowthAGaAAl2023} and for $\beta$-Ga$_2$O$_3$ homoepitaxy. For the latter, MEXCAT significantly widens the substrate-temperature ($T_\mathrm{sub}$) window, spanning from 700~°C to 900 °C depending on the growth orientation\cite{mazzoliniFacetingMetalexchangeCatalysis2019, mazzoliniSubstrateorientationDependenceVGa2020, mauzeMetalOxideCatalyzed2020, mazzolini_offcut-related_2020} [for (100), ($\bar{2}$01), (001), (010) from lowest to highest possible $T_\mathrm{sub}$]. In the homoepitaxial growths, despite providing a continuous flow of the catalyst no stabilization of the $\kappa$-phase was reported.\cite{mazzoliniSubstrateorientationDependenceVGa2020,mauzeMetalOxideCatalyzed2020} Apart from the possible role of the underlying substrate ($\beta$-Ga$_2$O$_3$ single crystal in the discussed case of homoepitaxy), the stabilization of $\beta$-Ga$_2$O$_3$ with MEXCAT could be related to the thermal meta-stability of the $\kappa$-phase. In fact, all the MEXCAT-MBE Ga$_2$O$_3$ depositions on different substrate orientations have been performed at $T_\mathrm{sub}$ $\geq$ 700 °C,\cite{mazzoliniSubstrateorientationDependenceVGa2020,mazzolini_offcut-related_2020,mauzeMetalOxideCatalyzed2020} while $\kappa$-Ga$_2$O$_3$ layers being their transformation to the thermodynamically stable $\beta$ phase upon ex-situ annealing at around 700 °C.\cite{fornariThermalStabilityEGa2O32017}\\

In this work, we use In-mediated MEXCAT in an O-plasma-assisted MBE system and identify the synthesis parameters that rule the stabilization of amorphous, $\beta$- and $\kappa$-Ga$_2$O$_3$ heteroepitaxial layers. The incorporation of In, leading to the formation of (In$_x$Ga$_{1-x}$)$_2$O$_3$ as well as impact of substrate on nucleation are investigated. Apart from providing practical guidelines for the MBE growth of phase pure $\kappa$/$\beta$-Ga$_2$O$_3$ and $\kappa$/$\beta$-(In$_x$Ga$_{1-x}$)$_2$O$_3$ layers, the collected experimental data demonstrate how MEXCAT can be used to significantly widen the $T_\mathrm{sub}$ window in $\beta$-Ga$_2$O$_3$ heteroepitaxy.

\section{\label{sec:Exp}Experiment}

Using In-mediated MEXCAT-MBE, we have grown epitaxial layers of Ga$_2$O$_3$ on $\alpha$-Al$_2$O$_3$ (0001) (i) with and (ii) without a $\beta$-Ga$_2$O$_3$ ($\bar{2}$01) nucleation layer "nl" containing rotational domains and (iii) single crystalline bulk $\beta$-Ga$_2$O$_3$ ($\bar{2}$01) substrates (unintentionally doped - Novel Crystal Technology, Inc.). Prior to the growth, the $\alpha$-Al$_2$O$_3$ (0001) substrates were solvent cleaned using acetone and isopropanol for 5 minutes in an ultrasonic bath, while the $\beta$-Ga$_2$O$_3$ ($\bar{2}$01) were chemically etched and annealed as described in a previous work.\cite{mazzoliniSubstrateorientationDependenceVGa2020} The substrates were all O$_{2}$ plasma treated inside the MBE chamber at a $T_\mathrm{sub}$ of 650 °C before deposition (\textit{$\Phi_{O_2}$} = 1 sccm, \textit{$P_\mathrm{RF}$} = 300 W, \textit{t} = 30 min). For all the growths on sapphire with "nl", a 20/30 nm thick $\beta$-Ga$_2$O$_3$ ($\bar{2}$01) layer was deposited at $T_\mathrm{sub}$ = 550 °C without supplying the additional In flux (\textit{i.e.}, conventional MBE). In-situ laser reflectometry (LR) and reflection high-energy electron diffraction (RHEED) were used to monitor the growth process. Elemental Ga and In were supplied by double filament effusion cells with a corresponding beam equivalent pressure (BEP, measured at the substrate position), of BEP$_{Ga}$ = 3.4 x 10$^{-7}$ mbar and BEP$_{In}$ = 1.2 x 10$^{-7}$ mbar for all the deposited samples. The fluxes were calibrated by measuring the growth rate of amorphous Ga$_2$O$_3$ (10.6 nm/min) and In$_2$O$_3$ (3.5 nm/min) layers at 100 °C, assuming full cation incorporation. A fixed oxygen flux \textit{$\Phi_{O_2}$} = 1 sccm was provided through an rf-plasma source (SPECS PCS). The impact of substrate temperature and oxidizing power on layer deposition was investigated by varying $T_\mathrm{sub}$ (measured through a thermocouple positioned on the backside of the substrate) between 540-760~°C and plasma power ($P_\mathrm{RF}$) between 145-225 W, respectively. 

To assess the crystal structure, x-ray diffraction measurements (XRD-PANalytical X’Pert Pro MRD) using the CuK$_{{\alpha}1}$ were performed on the grown layers in both symmetric out-of-plane (2$\theta$-$\omega$) and asymmetric reflections ($\Phi$-scan). The surface morphology was studied both with an atomic force microscope (AFM-Bruker Dimension Edge) in the PeakForce tapping mode and a scanning electron microscope (SEM). In order to locally investigate the crystal structure of morphological features highlighted by AFM, electron backscatter diffraction (EBSD-EDAX Hikari Super) has been performed on selected samples in the SEM. The incorporation of In during the MEXCAT growth was determined by time-of-flight secondary ion mass spectrometry (ToF-SIMS IV, iontof GmbH, Germany) using (In$_x$Ga$_{1-x}$)$_2$O$_3$ reference samples with known In-content as explained elsewhere.\cite{mazzoliniSubstrateorientationDependenceVGa2020} 

\section{\label{sec:Res}Results and discussion}

\subsection{\label{sec:Hetero}Phase and composition control on $\alpha$-Al$_2$O$_3$ (0001) with ($\bar{2}$01) $\beta$-Ga$_2$O$_3$ nl}

\begin{figure}
\includegraphics[width=1\linewidth]{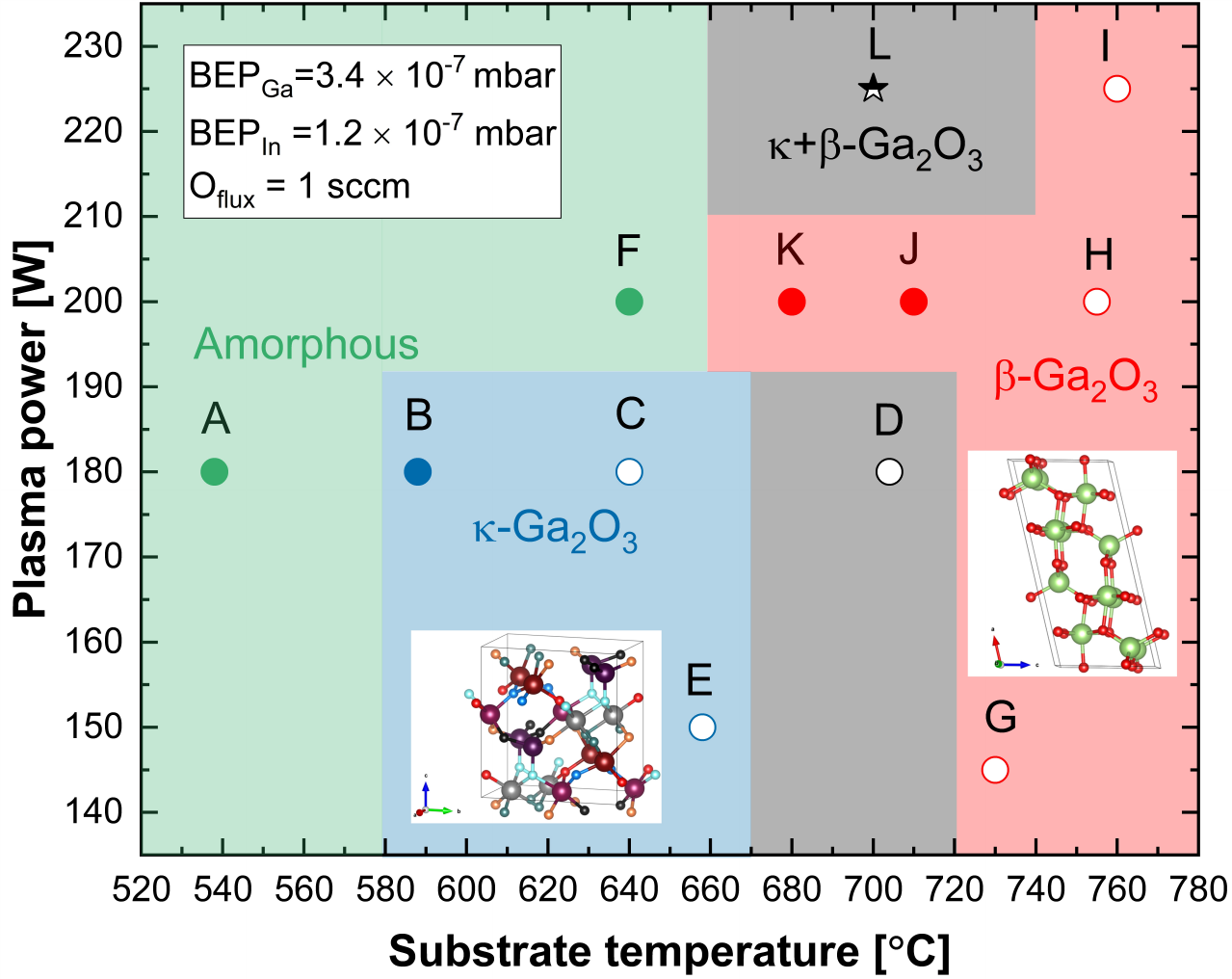}
\caption{\label{Fig.1} Phase diagram obtained from the MEXCAT-MBE growths as a function of plasma power and substrate temperature; filled symbols correspond to samples with In > 10\%, half-filled to In = 3\%, and open to In < 1\%. Sample L it's represented with a star because of possible nucleation issues (Supplementary). The sketches of the $\beta$- and $\kappa$-Ga$_2$O$_3$ unit cell were created using VESTA.\cite{mommaVESTAThreedimensionalVisualization2011}}
\end{figure}

In Table \ref{tab:table1}, we summarize the growth parameters for all the layers grown on $\alpha$-Al$_2$O$_3$ (0001) with ($\bar{2}$01) $\beta$-Ga$_2$O$_3$ "nl"; the same data are also summarized in Fig. \ref{Fig.1} highlighting the different phases and related In incorporation as a function of the $T_\mathrm{sub}$ and \textit{$P_\mathrm{RF}$} (constant metal and oxygen fluxes).

\begin{table}[t]
\caption{\label{tab:table1} Growth parameters of the heteroepitaxial samples presented in this work. All samples reported in this table were grown on a 20/30 nm $\beta$-Ga$_2$O$_3$ nucleation layer on $\alpha$-Al$_2$O$_3$ (0001).}
\begin{ruledtabular}
\begin{tabular}{ccccccc}
 Sample\footnote{BEP$_{Ga}$= 3.4 x 10$^{-7}$mbar, BEP$_{In}$= 1.2 x 10$^{-7}$mbar, O$_{flux}$ = 1 sccm for all samples} &$P_{rf}$ &$T_{sub}$ & In &Phase & G.R. & RMS\footnote{Calculated by 5x5 $\mu$m AFM images}\\
 & (W) & (°C) & (cat.\%) & & (nm/min)& (nm)\\
\hline
A& 180 & 540 & 15 & Amorphous & 9.8 & 26.8 \\
B& 180 & 590 & 14 & $\kappa$ & 9.5 & 2.2\\
C& 180 & 640 & 0.7 & $\kappa$ & 9 & 4.2 \\
D& 180 & 705 & 0.04 & $\kappa$+$\beta$ & 5.6 & 31.2\\
E& 150 & 660 & 0.02 & $\kappa$ & 6.3 & 13.1\\
F& 200 & 640 & 22 & Amorphous & 9.5 & 3.8\\
G& 145 & 730 & 0.02 & $\beta$ & 0.9 & 7.7\\
H& 200 & 755 & 0.02 & $\beta$ & 4.5 & 19.5\\
I& 225 & 760 & 0.02 & $\beta$ & 4.9 & 39.8\\
J& 200 & 710 & 8 & $\beta$ & 7.7 & 3.5\\
K& 200 & 680 & 17 & $\beta$ & 8.8 & 3.5\\
L& 225 & 700 & 3 & $\kappa$+$\beta$ & 9.6 & 35.0\\

\end{tabular}
\end{ruledtabular}
\end{table}

%\begin{table}[t]
%\caption{\label{tab:table1} Growth parameters of the heteroepitaxial samples presented in this work. All samples reported in this table were grown on a 20/30 nm $\beta$-Ga$_2$O$_3$ nucleation on $\alpha$-Al$_2$O$_3$ (0001).}
%\begin{ruledtabular}
%\begin{tabular}{ccccccc}
% Sample\footnote{BEP$_{Ga}$= 3.4 x 10$^{-7}$mbar, BEP$_{In}$= 1.2 x 10$^{-7}$mbar, O$_{flux}$ = 1 sccm for all samples} &$P_{rf}$ &$T_{sub}$ & In\% &Phase & G.R. & RMS\footnote{Calculated by 5x5 $\mu$m AFM images}\\
% & (W) & (°C) & (cat\%) & & (nm/min)& (nm)\\
%\hline
%A& 180 & 540 & 15 & Amorphous & 9.8 & 26.8 \\
%B& 200 & 640 & 22 & Amorphous & 9.5 & 3.8\\
%C& 200 & 680 & 17 & $\beta$ & 8.8 & 3.5\\
%D& 200 & 710 & 8 & $\beta$ & 7.7 & 3.5\\
%E& 180 & 590 & 14 & $\kappa$ & 9.5 & 2.2\\
%F& 180 & 640 & 0.7 & $\kappa$ & 9 & 4.2 \\
%G& 150 & 660 & 0.02 & $\kappa$ & 6.3 & 13.1\\
%H& 180 & 705 & 0.04 & $\kappa$+$\beta$ & 5.6 & 31.2\\
%I& 225 & 700 & 3 & $\kappa$+$\beta$ & 9.6 & 35.0\\
%J& 145 & 730 & 0.02 & $\beta$ & 0.9 & 7.7\\
%K& 225 & 760 & 0.02 & $\beta$ & 4.9 & 39.8\\
%L& 200 & 755 & 0.02 & $\beta$ & 4.5 & 19.5\\

%\end{tabular}
%\end{ruledtabular}
%\end{table}

\begin{figure*}[t]
\includegraphics[width=0.95\linewidth]{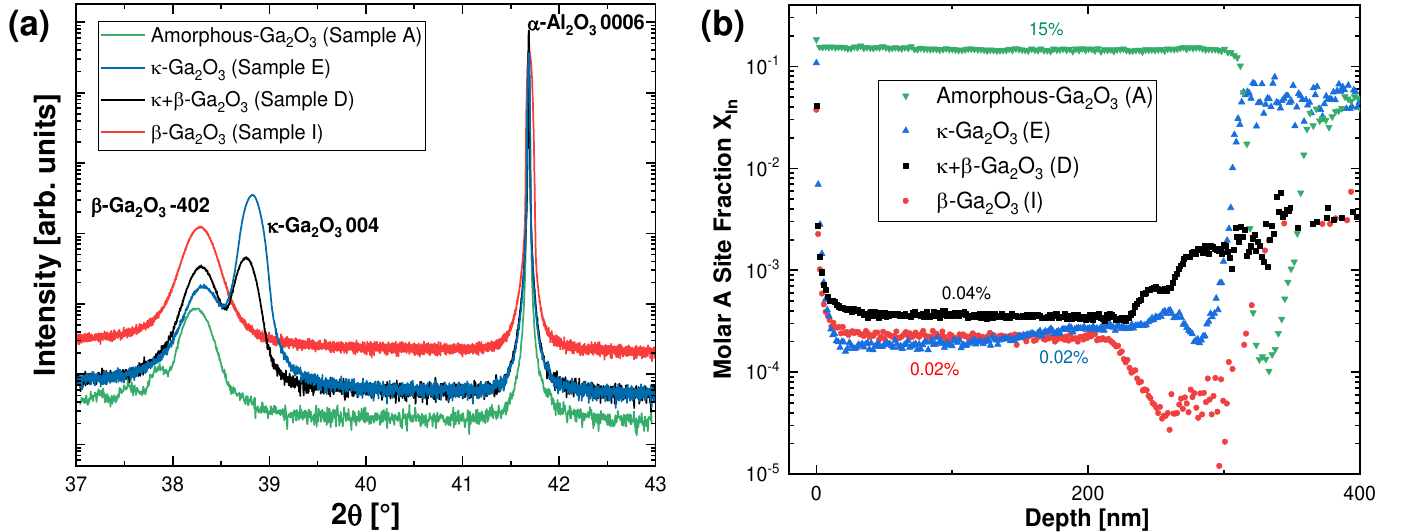}
\caption{\label{Fig.2} (a) Exemplary XRD 2$\theta$-$\omega$ scans highlight the difference between the different phases. (b) SIMS data for the In incorporation into the films/layers shown in (a).}
\end{figure*}

At low temperature ($T_\mathrm{sub}$= 540 °C) the growth resulted in amorphous material (sample A). While oscillations were recorded from LR (supplementary Fig. S1) highlighting the deposition of a layer, the RHEED pattern of the "nl" immediately disappeared with the beginning of the MEXCAT growth (supplementary Fig. S1). XRD data [2$\theta$-$\omega$ scan Fig. \ref{Fig.2}(a)] presents only the diffraction peak of the $\beta$-Ga$_2$O$_3$ polymorph, related to the "nl" previously deposited without catalysis. Notably, at this $T_\mathrm{sub}$, regardless of the $P_\mathrm{RF}$, epitaxial $\beta$-Ga$_2$O$_3$ can be deposited in conventional MBE, while the presence of the additional In-flux results in amorphous layers. The layer amorphization is probably related to the high amount of In incorporated under such synthesis conditions [around 15 cation percent - 15 cat.\% - see Fig. 2(b)] that increases the overall structural disorder.\cite{reisDisorderOrderTransition2023} The transition to the amorphous phase during MEXCAT is in agreement with findings on Sn-assisted growth for either low $T_\mathrm{sub}$\cite{kneissTinassistedHeteroepitaxialPLDgrowth2019} in PLD or too high catalyst flux in MBE.\cite{krachtTinAssistedSynthesisGa2017} Due to the enhanced disorder with increasing In content, the presence of micro-inclusions of $\gamma$-Ga$_2$O$_3$ in our layers is possible, which may not be visible with XRD. Furthermore, due to overlap of the $\gamma$-Ga$_2$O$_3$ 111 reflex and the $\beta$-Ga$_2$O$_3$ $\bar{2}$01 in 2$\theta$-$\omega$ scans,\cite{GammaDisorder} simple out-of plane XRD measurements might not be sufficient to exclude the presence of $\gamma$-inclusions.

\begin{figure*}[t]
\includegraphics[width=0.7\linewidth]{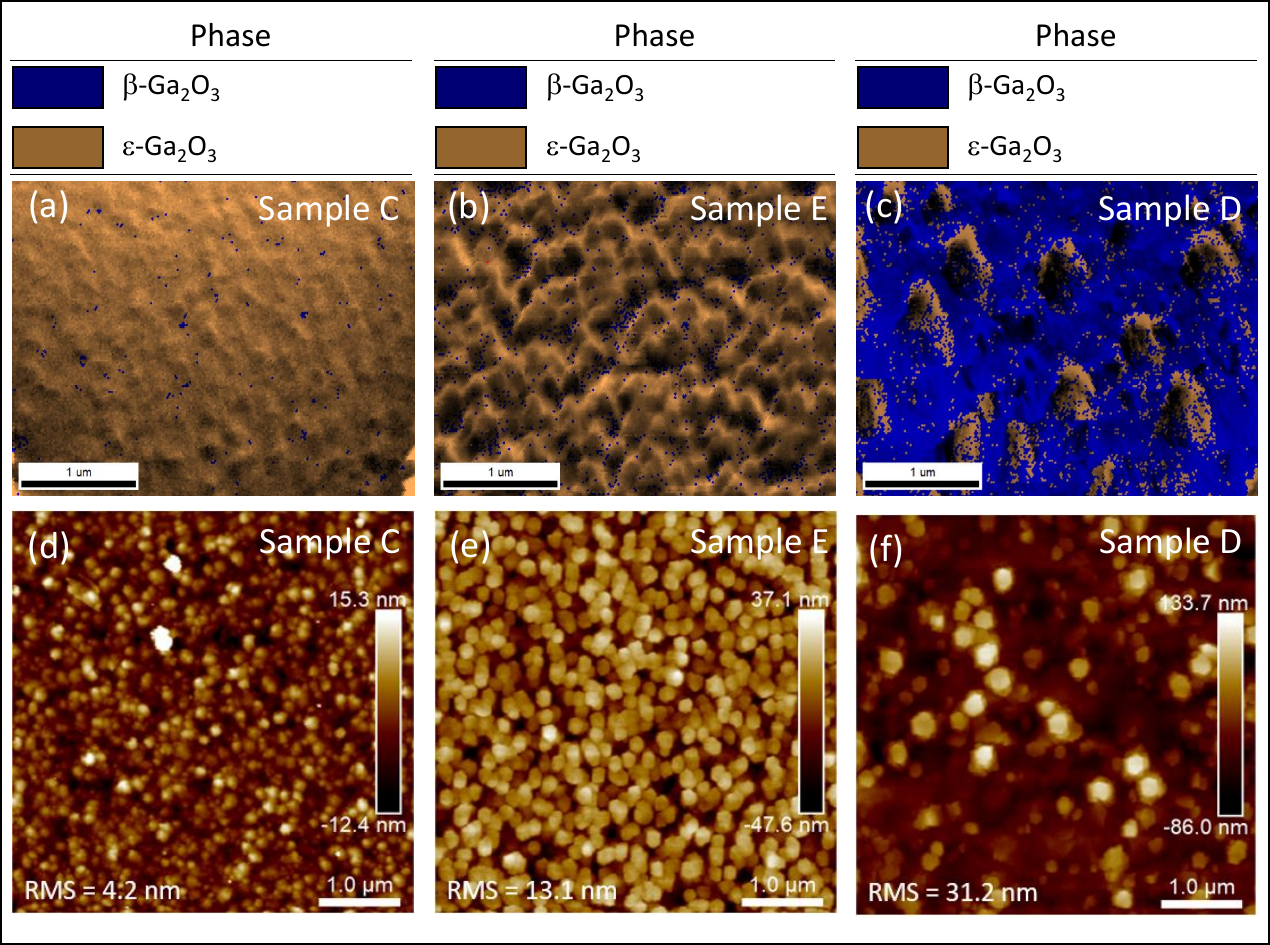}
\caption{\label{Fig.3} (a)-(c) EBSD phase maps (3 x 3.5 $\mu$m$^2$) superimposed on corresponding grayscale SEM images with their corresponding AFM image (d)-(f) (5 x 5 $\mu$m$^2$) for (a/d) sample C, (b/e) sample E, (c/f) sample D. As EBSD averages over the three microscopic in-plane domains of $\kappa$-Ga$_2$O$_3$, this phase is indexed as $\epsilon$-Ga$_2$O$_3$.}
\end{figure*}

The systematic effect of the $T_\mathrm{sub}$ on phase and In incorporation can be highlighted following the series of samples A, B, C, and D that were all deposited at the same $P_\mathrm{RF}$=180 W. In particular, an increase of 50 °C with respect to the growth temperature of sample A (sample B, $T_\mathrm{sub}$ = 590 °C) results in a phase-pure $\kappa$ layer which shows a decrease in the incorporated In content that is though still in the alloy regime [14 cat. \%, XRD characterization reported in supplementary Fig. S2].  In particular, the out-of-plane lattice constant of the $\kappa$-(In$_{0.14}$Ga$_{0.86}$)$_2$O$_3$ alloy increases to c = 9.4 Å (with respect to c = 9.283 Å expected for pure $\kappa$-Ga$_2$O$_3$).\cite{coraRealStructureEGa2017} Both the increase of the c lattice constant and the phase purity of the $\kappa$-layer, despite the high In concentration, are in good agreement with the work of Nishinaka et al. (phase separation observed just for In concentration $\geq$ 20\%).\cite{nishinakaIncorporationIndiumEgallium2018} On the other hand, an In content of 28\% can be incorporated using PLD while maintaining phase pure $\kappa$-Ga$_2$O$_3$.\cite{PLDupto28} A drastic reduction of In incorporation while keeping the pure orthorhombic phase can be obtained for a further increase of temperature (sample C, $T_\mathrm{sub}$ = 640 °C, In 0.7 \%) ending up with better crystal quality [full width at half maximum FWHM of 0.47° for the $\omega$-rocking curve of the 004 reflex, see supplementary Fig. S3]. The edge of the stability window of pure $\kappa$-Ga$_2$O$_3$ seems to be related to the phase transition temperature of this polymorph into the thermodynamically stable $\beta$-Ga$_2$O$_3$, in fact, sample D ($T_\mathrm{sub}$ = 705 °C) could be identified by both XRD [Fig. \ref{Fig.2}(a)] and EBSD [Fig. \ref{Fig.3}(c)] as a mixed $\kappa$+$\beta$ layer. 

The other parameter investigated in this work, the plasma power, is also affecting both the phase stabilization as well as the In incorporation. A decrease of plasma power with respect to sample C, for instance, allows to reduce the In concentration down to 0.02 cat\% while keeping the ortorhombic $\kappa$ phase (sample E, $P_\mathrm{RF}$ = 150 W, $T_\mathrm{sub}$ = 660 °C). The reduction of active oxygen thus favors the desorption of indium while maintaining a high Ga incorporation ($\sim$60 \%) while maintaining a similar crystal quality (FWHM = 0,52° - supplementary Fig. S3).

On the other hand, sample F ($P_\mathrm{RF}$ = 200 W, $T_\mathrm{sub}$ = 640°C) demonstrates how an increase of 20 W with respect to the conditions of the orthorhombic sample C drastically increases the In incorporation up to 22 cat.\%, resulting in an amorphous layer. This result suggests the presence of a $P_\mathrm{RF}$ upper threshold for the stability window of the $\kappa$ polymorph. As in the case of sample A, this could be related to the disorder induced by the large amount of incorporated In, suggesting that with MEXCAT-MBE the In incorporation in $\kappa$-(In$_x$Ga$_{1-x}$)$_2$O$_3$ layers could be limited to In concentration below 22 cat.\%, in good agreement with recent findings.\cite{IndiumSurfKarg} The trend of increased In-incorporation under higher oxidizing power (a higher plasma power used here or a higher oxygen flow rate used in Ref. \cite{hassaControlPhaseFormation2020}) or lower $T_\mathrm{sub}$ in MEXCAT-MBE observed here is in line with previously reported literature data.\cite{hassaControlPhaseFormation2020}

Additionally, the growth parameters in the $\kappa$ stability window have an influence on the obtained thin film morphology, which could have a strong influence on surface scattering in properly designed heterostructure (e.g., $\kappa$-AlGaO$_3$/$\kappa$-Ga$_2$O$_3$).\cite{rangaHighlyTunablePolarizationengineered2020}  We focus on the low In-content $\kappa$ layers C and E. Figure 3 shows EBSD phase maps for these in correlation to their surface morphology measured by AFM. The different plasma power [$P_\mathrm{RF}$ = 180 W for Fig. 3(a) and $P_\mathrm{RF}$ = 150 W for Fig. 3(b)] seems to have a significant impact on the overall morphology of the film. In the case of sample C, the layer is smoother (rms = 4.2 nm for a thickness of 504 nm) and homogeneous with respect to sample E that presents columnar islands with a well-defined hexagonal shape typical for the $\kappa$-Ga$_2$O$_3$ layers with 120°-rotated domains (in the past giving rise to the erroneous identification of a hexagonal $\epsilon$ phase).\cite{coraRealStructureEGa2017} It is reasonable to assume that the different $P_\mathrm{RF}$ could affect the diffusion length of adsorbed species eventually resulting in different layer morphologies. The rotational domains in $\kappa$ layers (expected average domains size in the range of 5-20 nm\cite{mazzoliniSilaneMediatedExpansion2023, coraRealStructureEGa2017}) are also related to the erroneously indexed $\epsilon$ phase by EBSD: since its lateral sensitivity is limited to about 20 nm, the technique is most likely considering the contribution from all of the three rotated domains of the orthorhombic symmetry ending up in a misleading labelling of the $\kappa$ phase.

After having identified the stability window of the orthorhombic Ga$_2$O$_3$ and (In,Ga)$_2$O$_3$ polymorph, the focus is shifted to the monoclinic system. Sample G is the layer deposited at the lowest plasma power at substrate temperature exceeding the stability window of the $\kappa$ phase ($P_\mathrm{RF}$ = 145 W, $T_\mathrm{sub}$ = 730 °C). As expected from previously discussed experimental findings, the sample is pure $\beta$-Ga$_2$O$_3$ with very limited In incorporation (0.02 cat.\%). Nonetheless, in such synthesis conditions the incorporation of the provided Ga flux is limited to less than 10\%. To enhance the growth rate (i.e., Ga incorporation), the $P_\mathrm{RF}$ can be increased. Samples H and I illustrate that at $P_\mathrm{RF}$ values of 200 and 225 W, the single $\beta$ phase can be maintained while simultaneously retaining the low
In content of sample G (0.02 cat.\%) due to the sufficiently high $T_\mathrm{sub}$ (760 and 755°C respectively for H and I) with Ga incorporation approaching 50\% of the provided Ga-flux.
The $\beta$-(In$_x$Ga$_{1-x}$)$_2$O$_3$ alloyed layers can be obtained for a sufficiently high plasma power ($P_\mathrm{RF}$ = 200 W) while the $T_\mathrm{sub}$ is decreased towards the stability window of the $\kappa$ polymorph (i.e., 710 and 680°C for sample J and K respectively): particularly, phase pure monoclinic alloyed layers up to 17 cat.\%  (sample K) were here demonstrated. 
Therefore, in the synthesis parameters space investigated in this work both the $\kappa$ and $\beta$-(In$_x$Ga$_{1-x}$)$_2$O$_3$ alloy material systems obtained via In-mediated MEXCAT-MBE seems to be limited to a similar composition range approaching the 15-20 cat.\%.

In this rather consistent framework, sample L seems
to be out of trend being probably mixed $\beta$/$\kappa$ phase and resulting in lower In incorporation with respect to samples K and J, even if being deposited at higher plasma power; we consider the presence of non-homogeneous nucleation as a possible explanation. A more detailed discussion concerning a possible non-homogeneous nucleation layer for sample L, as well as more insight on the mixed $\kappa$/$\beta$ phase are provided in the supplementary material.

\subsection{\label{sec:Homo} Nucleation of $\kappa$-Ga$_2$O$_3$ on different substrates}

\begin{figure}
\includegraphics[width=0.88\linewidth]{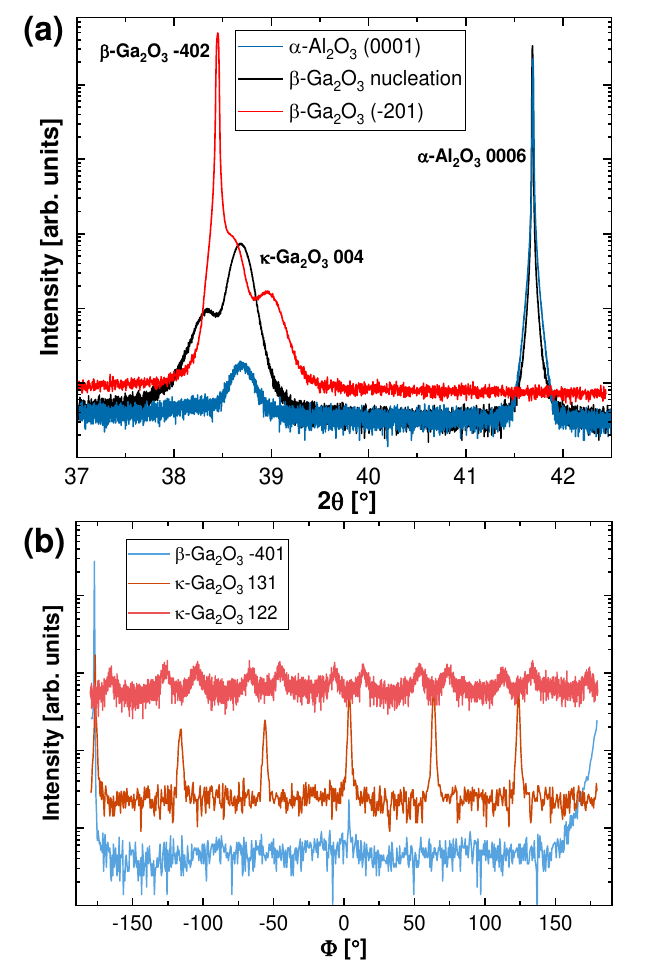}
\caption{\label{fig:4} (a) XRD 2$\theta$-$\omega$ scans for the 3 different co-loaded substrates, for all of them we can see the presence of the k-Ga$_2$O$_3$ 004 reflex. (b) $\Phi$-scan for the 202 and 131 reflexes of the $\kappa$-Ga$_2$O$_3$ and the $\bar{4}$01 for the $\beta$-Ga$_2$O$_3$ for the sample grown on $\beta$-Ga$_2$O$_3$ ($\bar{2}$01).}
\end{figure}

The growth on top of single crystalline bulk ($\bar{2}$01)-oriented $\beta$-Ga$_2$O$_3$ substrate could  be a different system for the $\kappa$-Ga$_2$O$_3$ stabilization with respect to the deposition on $\alpha$-Al$_2$O$_3$ (0001) + $\beta$ ($\bar{2}$01) nucleation layer. This is related to the strain and structural defects (including rotational domains) which are intrinsically present in the $\beta$ heteroepitaxial nl on sapphire\cite{https://doi.org/10.1002/pssa.201329040,PhysRevMaterials.4.104602,rafique_towards_2018} and that could affect the nucleation of the orthorhombic polymorph and its structural quality. Other growth techniques, such as halide vapor phase epitaxy (HVPE) already demonstrated the possibility to stabilize the $\kappa$ phase with rotational domains on $\beta$-Ga$_2$O$_3$ ($\bar{2}$01) substrates;\cite{oshimaEpitaxialGrowthPhasepure2015} nonetheless, to the best of our knowledge, there is no report of MBE grown layers of $\kappa$-Ga$_2$O$_3$ on ($\bar{2}$01) $\beta$-Ga$_2$O$_3$ bulk substrates. For this growth, three different substrates where co-loaded, a $\beta$-Ga$_2$O$_3$ ($\bar{2}$01) single crystal, a $\alpha$-Al$_2$O$_3$ (0001) with a 70/80 nm $\beta$-Ga$_2$O$_3$ ($\bar{2}$01) nucleation layer on top (conventional MBE growth performed in a previous deposition run) and a bare $\alpha$-Al$_2$O$_3$ (0001). The growth was performed directly with MEXCAT with the very same growth conditions adopted for sample E, that resulted in pure $\kappa$-Ga$_2$O$_3$. After the deposition, all the substrates showed the presence of the $\kappa$-phase [Fig. \ref{fig:4}(a)] even if a clear difference can be highlighted. As expected the sample with a nucleation layer of 70/80 nm $\beta$-Ga$_2$O$_3$ reproduced the result already seen for sample E. Direct deposition on $\alpha$-Al$_2$O$_3$ (0001) still resulted in the orthorhombic phase as highlighted by the 004 reflex from the symmetric on-axis 2$\theta$-$\omega$ scan in Fig. \ref{fig:4}(a) but with a significantly lower intensity compared to the samples grown on the $\beta$ nucleation layer. This is probably related to the impeded nucleation of $\kappa$ directly on the bare $\alpha$-Al$_2$O$_3$ (0001) substrate.\\
\indent Instead, for the growth on the bulk ($\bar{2}$01) $\beta$-Ga$_2$O$_3$ substrate two different peaks at 2$\theta$ = 38.63° and 2$\theta$ = 38.96° are visible Fig. \ref{fig:4}(a). The peak at lower 2$\theta$ angle is arising from the $\beta$-Ga$_2$O$_3$ bulk substrate (supplementary Fig. S4). The reflex at higher 2$\theta$ angle (38.96°) instead seems to be similar to the results of Nishinaka \textit{et. al.}\cite{doi:10.1021/acsomega.0c04634} for the growth of single crystal $\kappa$-Ga$_2$O$_3$ on $\epsilon$-GaFeO$_3$. From rocking curve values around the 004 reflex (FWHM = 0.68° - supplementary Fig. S5) the crystal quality of the $\kappa$-Ga$_2$O$_3$ (001) layer grown on $\beta$-Ga$_2$O$_3$ ($\bar{2}$01) single crystal seems to slightly increase in comparison to the growth on c-Al$_2$O$_3$ with nucleation (FWHM = 0.78° - supplementary Fig. S5), but is still not compatible with a $\kappa$-Ga$_2$O$_3$ single crystal\cite{doi:10.1021/acsomega.0c04634} nor with a sensible enlargement of the overall domain size.\cite{mazzoliniSilaneMediatedExpansion2023} Nevertheless, as also shown from $\Phi$-scans for the $\kappa$ 131 and 122 reflections [Fig. \ref{fig:4}(b)], $\kappa$-Ga$_2$O$_3$ can be stabilized also on ($\bar{2}$01) Ga$_2$O$_3$ bulk substrates with MEXCAT-MBE. Figure \ref{fig:4}(b) shows that rotational domains are still present in the $\kappa$-Ga$_2$O$_3$ grown layer, as already seen for halide vapor phase epitaxy.\cite{oshimaEpitaxialGrowthPhasepure2015}

Our study suggests that the competing stabilization of the $\kappa$ and $\beta$ polymorphs in MEXCAT is strongly affected by the thermal stability of the metastable $\kappa$ polymorph. Nevertheless, additional experiments may be performed in order to exclude the role of strain and defects on the polymorph stabilization (\textit{e.g}, different $\beta$-Ga$_2$O$_3$ substrate orientations).

\section{\label{sec:Conc}Conclusions}
\nocite{*}

In summary, we have used In-mediated MEXCAT in a PAMBE system in order to identify the synthesis parameters (substrate temperature $T_\mathrm{sub}$ and oxidizing power controlled by RF plasma power $P_\mathrm{RF}$) that result in the formation of amorphous, $\beta$ and $\kappa$-Ga$_2$O$_3$ as well as (In$_x$Ga$_{1-x}$)$_2$O$_3$ heteroepitaxial layers on $\alpha$-Al$_2$O$_3$ (0001) with a ($\bar{2}$01) $\beta$-Ga$_2$O$_3$ nucleation layer.
The deposition of $\kappa$ layers can be obtained at $T_\mathrm{sub}$ below its polymorph stability limit (700 °C) at sufficiently low $P_\mathrm{RF}$. Increasing oxidizing power or decreasing $T_\mathrm{sub}$ increases the In-incorporation, eventually inducing sufficient structural disorder to form amorphous layers. As a result, In can be incorporated in the $\kappa$-(In$_x$Ga$_{1-x}$)$_2$O$_3$ alloy system at least up to 14 cat.\% without formation of amorphous layer for $T_\mathrm{sub}$ $\geq$ 580 °C. 
The heteroepitaxial deposition of the thermodynamically stable $\beta$-Ga$_2$O$_3$ layers can be obtained with MEXCAT for $T_\mathrm{sub}$ $\geq$ 730 °C for various plasma powers, significantly widening its growth window by at least 100 °C with respect to conventional non-catalyzed MBE. $\beta$-(In$_x$Ga$_{1-x}$)$_2$O$_3$ layers with In concentrations as high as $x$ = 0.17 can be obtained when lowering the $T_\mathrm{sub}$ at sufficiently high plasma powers.
We have thoroughly discussed how the growth parameters can affect the morphology and the crystal quality of the deposited $\kappa$-phase layers. 
The comparison between the growth on defective $\beta$-Ga$_2$O$_3$ nucleation layers, single crystalline bulk $\beta$ substrates with the same ($\bar{2}$01) orientation, and pure sapphire substrates, suggests a negligible role of strain/structural defects on the $\kappa$ phase stabilization by MEXCAT-MBE, but the importance of underlying Ga$_2$O$_3$ for layer nucleation.

\section*{Supplementary material}

A more detailed investigation concerning a possible
non-homogeneous nucleation layer for sample H and I, as well as more insight on the mixed $\kappa$/$\beta$ phase can be found in the supplementary material. AFM 5x5 $\mu$m$^2$ images and 2$\theta$-$\omega$ wide scan for all the layer grown on $\alpha$-Al$_2$O$_3$ (0001) with $\beta$-Ga$_2$O$_3$ ($\bar{2}$01) nl can also be found.

\section*{Acknowledgments}
We would like to thank Philipp John for critically reading the manuscript, as well as Hans-Peter Sch{\"o}nherr, Claudia Herrmann, Carsten Stemmler, and Steffen Behnke for their
technical support on the MBE. This work was
performed in the framework of GraFOx, a Leibniz-ScienceCampus, and was funded by Deutsche Forschungsgemeinschaft (DFG,
German Research Foundation)—Project No. 446185170.

\section*{Authors declarations}
\subsection*{Conflict of interest}
The authors have no conflicts to declare.

\section*{Data availability statement}

The data that support the findings of this study are available within the article and its Supplementary Material. Additional data may be obtained from the corresponding authors upon reasonable request.

\section*{References}
\bibliography{main}% Produces the bibliography via BibTeX.

\begin{thebibliography}{45}%
\makeatletter
\providecommand \@ifxundefined [1]{%
 \@ifx{#1\undefined}
}%
\providecommand \@ifnum [1]{%
 \ifnum #1\expandafter \@firstoftwo
 \else \expandafter \@secondoftwo
 \fi
}%
\providecommand \@ifx [1]{%
 \ifx #1\expandafter \@firstoftwo
 \else \expandafter \@secondoftwo
 \fi
}%
\providecommand \natexlab [1]{#1}%
\providecommand \enquote  [1]{``#1''}%
\providecommand \bibnamefont  [1]{#1}%
\providecommand \bibfnamefont [1]{#1}%
\providecommand \citenamefont [1]{#1}%
\providecommand \href@noop [0]{\@secondoftwo}%
\providecommand \href [0]{\begingroup \@sanitize@url \@href}%
\providecommand \@href[1]{\@@startlink{#1}\@@href}%
\providecommand \@@href[1]{\endgroup#1\@@endlink}%
\providecommand \@sanitize@url [0]{\catcode `\\12\catcode `\$12\catcode
  `\&12\catcode `\#12\catcode `\^12\catcode `\_12\catcode `\%12\relax}%
\providecommand \@@startlink[1]{}%
\providecommand \@@endlink[0]{}%
\providecommand \url  [0]{\begingroup\@sanitize@url \@url }%
\providecommand \@url [1]{\endgroup\@href {#1}{\urlprefix }}%
\providecommand \urlprefix  [0]{URL }%
\providecommand \Eprint [0]{\href }%
\providecommand \doibase [0]{http://dx.doi.org/}%
\providecommand \selectlanguage [0]{\@gobble}%
\providecommand \bibinfo  [0]{\@secondoftwo}%
\providecommand \bibfield  [0]{\@secondoftwo}%
\providecommand \translation [1]{[#1]}%
\providecommand \BibitemOpen [0]{}%
\providecommand \bibitemStop [0]{}%
\providecommand \bibitemNoStop [0]{.\EOS\space}%
\providecommand \EOS [0]{\spacefactor3000\relax}%
\providecommand \BibitemShut  [1]{\csname bibitem#1\endcsname}%
\let\auto@bib@innerbib\@empty
%</preamble>
\bibitem [{\citenamefont {Green}\ \emph {et~al.}(2022)\citenamefont {Green},
  \citenamefont {Speck}, \citenamefont {Xing}, \citenamefont {Moens},
  \citenamefont {Allerstam}, \citenamefont {Gumaelius}, \citenamefont {Neyer},
  \citenamefont {Arias-Purdue}, \citenamefont {Mehrotra}, \citenamefont
  {Kuramata}, \citenamefont {Sasaki}, \citenamefont {Watanabe}, \citenamefont
  {Koshi}, \citenamefont {Blevins}, \citenamefont {Bierwagen}, \citenamefont
  {Krishnamoorthy}, \citenamefont {Leedy}, \citenamefont {Arehart},
  \citenamefont {Neal}, \citenamefont {Mou}, \citenamefont {Ringel},
  \citenamefont {Kumar}, \citenamefont {Sharma}, \citenamefont {Ghosh},
  \citenamefont {Singisetti}, \citenamefont {Li}, \citenamefont {Chabak},
  \citenamefont {Liddy}, \citenamefont {Islam}, \citenamefont {Rajan},
  \citenamefont {Graham}, \citenamefont {Choi}, \citenamefont {Cheng},\ and\
  \citenamefont {Higashiwaki}}]{doi:10.1063/5.0060327}%
  \BibitemOpen
  \bibfield  {author} {\bibinfo {author} {\bibfnamefont {A.~J.}\ \bibnamefont
  {Green}}, \bibinfo {author} {\bibfnamefont {J.}~\bibnamefont {Speck}},
  \bibinfo {author} {\bibfnamefont {G.}~\bibnamefont {Xing}}, \bibinfo {author}
  {\bibfnamefont {P.}~\bibnamefont {Moens}}, \bibinfo {author} {\bibfnamefont
  {F.}~\bibnamefont {Allerstam}}, \bibinfo {author} {\bibfnamefont
  {K.}~\bibnamefont {Gumaelius}}, \bibinfo {author} {\bibfnamefont
  {T.}~\bibnamefont {Neyer}}, \bibinfo {author} {\bibfnamefont
  {A.}~\bibnamefont {Arias-Purdue}}, \bibinfo {author} {\bibfnamefont
  {V.}~\bibnamefont {Mehrotra}}, \bibinfo {author} {\bibfnamefont
  {A.}~\bibnamefont {Kuramata}}, \bibinfo {author} {\bibfnamefont
  {K.}~\bibnamefont {Sasaki}}, \bibinfo {author} {\bibfnamefont
  {S.}~\bibnamefont {Watanabe}}, \bibinfo {author} {\bibfnamefont
  {K.}~\bibnamefont {Koshi}}, \bibinfo {author} {\bibfnamefont
  {J.}~\bibnamefont {Blevins}}, \bibinfo {author} {\bibfnamefont
  {O.}~\bibnamefont {Bierwagen}}, \bibinfo {author} {\bibfnamefont
  {S.}~\bibnamefont {Krishnamoorthy}}, \bibinfo {author} {\bibfnamefont
  {K.}~\bibnamefont {Leedy}}, \bibinfo {author} {\bibfnamefont {A.~R.}\
  \bibnamefont {Arehart}}, \bibinfo {author} {\bibfnamefont {A.~T.}\
  \bibnamefont {Neal}}, \bibinfo {author} {\bibfnamefont {S.}~\bibnamefont
  {Mou}}, \bibinfo {author} {\bibfnamefont {S.~A.}\ \bibnamefont {Ringel}},
  \bibinfo {author} {\bibfnamefont {A.}~\bibnamefont {Kumar}}, \bibinfo
  {author} {\bibfnamefont {A.}~\bibnamefont {Sharma}}, \bibinfo {author}
  {\bibfnamefont {K.}~\bibnamefont {Ghosh}}, \bibinfo {author} {\bibfnamefont
  {U.}~\bibnamefont {Singisetti}}, \bibinfo {author} {\bibfnamefont
  {W.}~\bibnamefont {Li}}, \bibinfo {author} {\bibfnamefont {K.}~\bibnamefont
  {Chabak}}, \bibinfo {author} {\bibfnamefont {K.}~\bibnamefont {Liddy}},
  \bibinfo {author} {\bibfnamefont {A.}~\bibnamefont {Islam}}, \bibinfo
  {author} {\bibfnamefont {S.}~\bibnamefont {Rajan}}, \bibinfo {author}
  {\bibfnamefont {S.}~\bibnamefont {Graham}}, \bibinfo {author} {\bibfnamefont
  {S.}~\bibnamefont {Choi}}, \bibinfo {author} {\bibfnamefont {Z.}~\bibnamefont
  {Cheng}}, \ and\ \bibinfo {author} {\bibfnamefont {M.}~\bibnamefont
  {Higashiwaki}},\ }\href@noop {} {\bibfield  {journal} {\bibinfo  {journal}
  {APL Materials}\ }\textbf {\bibinfo {volume} {10}} (\bibinfo {year}
  {2022})}\BibitemShut {NoStop}%
\bibitem [{\citenamefont {Galazka}\ \emph {et~al.}(2017)\citenamefont
  {Galazka}, \citenamefont {Uecker}, \citenamefont {Klimm}, \citenamefont
  {Irmscher}, \citenamefont {Naumann}, \citenamefont {Pietsch}, \citenamefont
  {Kwasniewski}, \citenamefont {Bertram}, \citenamefont {Ganschow},\ and\
  \citenamefont {Bickermann}}]{galazkaScalingUpBulkVGa2017}%
  \BibitemOpen
  \bibfield  {author} {\bibinfo {author} {\bibfnamefont {Z.}~\bibnamefont
  {Galazka}}, \bibinfo {author} {\bibfnamefont {R.}~\bibnamefont {Uecker}},
  \bibinfo {author} {\bibfnamefont {D.}~\bibnamefont {Klimm}}, \bibinfo
  {author} {\bibfnamefont {K.}~\bibnamefont {Irmscher}}, \bibinfo {author}
  {\bibfnamefont {M.}~\bibnamefont {Naumann}}, \bibinfo {author} {\bibfnamefont
  {M.}~\bibnamefont {Pietsch}}, \bibinfo {author} {\bibfnamefont
  {A.}~\bibnamefont {Kwasniewski}}, \bibinfo {author} {\bibfnamefont
  {R.}~\bibnamefont {Bertram}}, \bibinfo {author} {\bibfnamefont
  {S.}~\bibnamefont {Ganschow}}, \ and\ \bibinfo {author} {\bibfnamefont
  {M.}~\bibnamefont {Bickermann}},\ }\href@noop {} {\bibfield  {journal}
  {\bibinfo  {journal} {ECS J. Solid State Sci. Technol.}\ }\textbf {\bibinfo
  {volume} {6}} (\bibinfo {year} {2017})}\BibitemShut {NoStop}%
\bibitem [{\citenamefont {Kuramata}\ \emph {et~al.}(2016)\citenamefont
  {Kuramata}, \citenamefont {Koshi}, \citenamefont {Watanabe}, \citenamefont
  {Yamaoka}, \citenamefont {Masui},\ and\ \citenamefont
  {Yamakoshi}}]{kuramataHighqualityVGaSingle2016a}%
  \BibitemOpen
  \bibfield  {author} {\bibinfo {author} {\bibfnamefont {A.}~\bibnamefont
  {Kuramata}}, \bibinfo {author} {\bibfnamefont {K.}~\bibnamefont {Koshi}},
  \bibinfo {author} {\bibfnamefont {S.}~\bibnamefont {Watanabe}}, \bibinfo
  {author} {\bibfnamefont {Y.}~\bibnamefont {Yamaoka}}, \bibinfo {author}
  {\bibfnamefont {T.}~\bibnamefont {Masui}}, \ and\ \bibinfo {author}
  {\bibfnamefont {S.}~\bibnamefont {Yamakoshi}},\ }\href@noop {} {\bibfield
  {journal} {\bibinfo  {journal} {Jpn. J. Appl. Phys.}\ }\textbf {\bibinfo
  {volume} {55}} (\bibinfo {year} {2016})}\BibitemShut {NoStop}%
\bibitem [{\citenamefont {Ohba}\ \emph {et~al.}(2021)\citenamefont {Ohba},
  \citenamefont {Kobayashi}, \citenamefont {Taishi},\ and\ \citenamefont
  {Hoshikawa}}]{ohbaGrowthVGa2O3Single2021a}%
  \BibitemOpen
  \bibfield  {author} {\bibinfo {author} {\bibfnamefont {E.}~\bibnamefont
  {Ohba}}, \bibinfo {author} {\bibfnamefont {T.}~\bibnamefont {Kobayashi}},
  \bibinfo {author} {\bibfnamefont {T.}~\bibnamefont {Taishi}}, \ and\ \bibinfo
  {author} {\bibfnamefont {K.}~\bibnamefont {Hoshikawa}},\ }\href@noop {}
  {\bibfield  {journal} {\bibinfo  {journal} {Journal of Crystal Growth}\
  }\textbf {\bibinfo {volume} {556}} (\bibinfo {year} {2021})}\BibitemShut
  {NoStop}%
\bibitem [{\citenamefont {V{\'i}llora}\ \emph {et~al.}(2004)\citenamefont
  {V{\'i}llora}, \citenamefont {Shimamura}, \citenamefont {Yoshikawa},
  \citenamefont {Aoki},\ and\ \citenamefont
  {Ichinose}}]{villoraLargesizeVGa2O3Single2004}%
  \BibitemOpen
  \bibfield  {author} {\bibinfo {author} {\bibfnamefont {E.~G.}\ \bibnamefont
  {V{\'i}llora}}, \bibinfo {author} {\bibfnamefont {K.}~\bibnamefont
  {Shimamura}}, \bibinfo {author} {\bibfnamefont {Y.}~\bibnamefont
  {Yoshikawa}}, \bibinfo {author} {\bibfnamefont {K.}~\bibnamefont {Aoki}}, \
  and\ \bibinfo {author} {\bibfnamefont {N.}~\bibnamefont {Ichinose}},\
  }\href@noop {} {\bibfield  {journal} {\bibinfo  {journal} {Journal of Crystal
  Growth}\ }\textbf {\bibinfo {volume} {270}} (\bibinfo {year}
  {2004})}\BibitemShut {NoStop}%
\bibitem [{\citenamefont {Mauze}\ \emph
  {et~al.}(2020{\natexlab{a}})\citenamefont {Mauze}, \citenamefont {Zhang},
  \citenamefont {Itoh}, \citenamefont {Ahmadi},\ and\ \citenamefont
  {Speck}}]{doi:10.1063/5.0027870}%
  \BibitemOpen
  \bibfield  {author} {\bibinfo {author} {\bibfnamefont {A.}~\bibnamefont
  {Mauze}}, \bibinfo {author} {\bibfnamefont {Y.}~\bibnamefont {Zhang}},
  \bibinfo {author} {\bibfnamefont {T.}~\bibnamefont {Itoh}}, \bibinfo {author}
  {\bibfnamefont {E.}~\bibnamefont {Ahmadi}}, \ and\ \bibinfo {author}
  {\bibfnamefont {J.~S.}\ \bibnamefont {Speck}},\ }\href@noop {} {\bibfield
  {journal} {\bibinfo  {journal} {Applied Physics Letters}\ }\textbf {\bibinfo
  {volume} {117}} (\bibinfo {year} {2020}{\natexlab{a}})}\BibitemShut {NoStop}%
\bibitem [{\citenamefont {Ahmadi}\ \emph {et~al.}(2017)\citenamefont {Ahmadi},
  \citenamefont {Koksaldi}, \citenamefont {Kaun}, \citenamefont {Oshima},
  \citenamefont {Short}, \citenamefont {Mishra},\ and\ \citenamefont
  {Speck}}]{Ahmadi_2017}%
  \BibitemOpen
  \bibfield  {author} {\bibinfo {author} {\bibfnamefont {E.}~\bibnamefont
  {Ahmadi}}, \bibinfo {author} {\bibfnamefont {O.~S.}\ \bibnamefont
  {Koksaldi}}, \bibinfo {author} {\bibfnamefont {S.~W.}\ \bibnamefont {Kaun}},
  \bibinfo {author} {\bibfnamefont {Y.}~\bibnamefont {Oshima}}, \bibinfo
  {author} {\bibfnamefont {D.~B.}\ \bibnamefont {Short}}, \bibinfo {author}
  {\bibfnamefont {U.~K.}\ \bibnamefont {Mishra}}, \ and\ \bibinfo {author}
  {\bibfnamefont {J.~S.}\ \bibnamefont {Speck}},\ }\href@noop {} {\bibfield
  {journal} {\bibinfo  {journal} {Applied Physics Express}\ }\textbf {\bibinfo
  {volume} {10}} (\bibinfo {year} {2017})}\BibitemShut {NoStop}%
\bibitem [{\citenamefont {Ardenghi}\ \emph {et~al.}(2022)\citenamefont
  {Ardenghi}, \citenamefont {Bierwagen}, \citenamefont {Falkenstein},
  \citenamefont {Hoffmann}, \citenamefont {Lähnemann}, \citenamefont
  {Martin},\ and\ \citenamefont {Mazzolini}}]{doi:10.1063/5.0087987}%
  \BibitemOpen
  \bibfield  {author} {\bibinfo {author} {\bibfnamefont {A.}~\bibnamefont
  {Ardenghi}}, \bibinfo {author} {\bibfnamefont {O.}~\bibnamefont {Bierwagen}},
  \bibinfo {author} {\bibfnamefont {A.}~\bibnamefont {Falkenstein}}, \bibinfo
  {author} {\bibfnamefont {G.}~\bibnamefont {Hoffmann}}, \bibinfo {author}
  {\bibfnamefont {J.}~\bibnamefont {Lähnemann}}, \bibinfo {author}
  {\bibfnamefont {M.}~\bibnamefont {Martin}}, \ and\ \bibinfo {author}
  {\bibfnamefont {P.}~\bibnamefont {Mazzolini}},\ }\href@noop {} {\bibfield
  {journal} {\bibinfo  {journal} {Applied Physics Letters}\ }\textbf {\bibinfo
  {volume} {121}} (\bibinfo {year} {2022})}\BibitemShut {NoStop}%
\bibitem [{\citenamefont {McCandless}\ \emph {et~al.}(2022)\citenamefont
  {McCandless}, \citenamefont {Protasenko}, \citenamefont {Morell},
  \citenamefont {Steinbrunner}, \citenamefont {Neal}, \citenamefont {Tanen},
  \citenamefont {Cho}, \citenamefont {Asel}, \citenamefont {Mou}, \citenamefont
  {Vogt}, \citenamefont {Xing},\ and\ \citenamefont
  {Jena}}]{doi:10.1063/5.0101132}%
  \BibitemOpen
  \bibfield  {author} {\bibinfo {author} {\bibfnamefont {J.~P.}\ \bibnamefont
  {McCandless}}, \bibinfo {author} {\bibfnamefont {V.}~\bibnamefont
  {Protasenko}}, \bibinfo {author} {\bibfnamefont {B.~W.}\ \bibnamefont
  {Morell}}, \bibinfo {author} {\bibfnamefont {E.}~\bibnamefont
  {Steinbrunner}}, \bibinfo {author} {\bibfnamefont {A.~T.}\ \bibnamefont
  {Neal}}, \bibinfo {author} {\bibfnamefont {N.}~\bibnamefont {Tanen}},
  \bibinfo {author} {\bibfnamefont {Y.}~\bibnamefont {Cho}}, \bibinfo {author}
  {\bibfnamefont {T.~J.}\ \bibnamefont {Asel}}, \bibinfo {author}
  {\bibfnamefont {S.}~\bibnamefont {Mou}}, \bibinfo {author} {\bibfnamefont
  {P.}~\bibnamefont {Vogt}}, \bibinfo {author} {\bibfnamefont {H.~G.}\
  \bibnamefont {Xing}}, \ and\ \bibinfo {author} {\bibfnamefont
  {D.}~\bibnamefont {Jena}},\ }\href@noop {} {\bibfield  {journal} {\bibinfo
  {journal} {Applied Physics Letters}\ }\textbf {\bibinfo {volume} {121}}
  (\bibinfo {year} {2022})}\BibitemShut {NoStop}%
\bibitem [{\citenamefont {Azizie}\ \emph {et~al.}(2023)\citenamefont {Azizie},
  \citenamefont {Hensling}, \citenamefont {Gorsak}, \citenamefont {Kim},
  \citenamefont {Pieczulewski}, \citenamefont {Dryden}, \citenamefont
  {Senevirathna}, \citenamefont {Coye}, \citenamefont {Shang}, \citenamefont
  {Steele}, \citenamefont {Vogt}, \citenamefont {Parker}, \citenamefont
  {Birkhölzer}, \citenamefont {McCandless}, \citenamefont {Jena},
  \citenamefont {Xing}, \citenamefont {Liu}, \citenamefont {Williams},
  \citenamefont {Green}, \citenamefont {Chabak}, \citenamefont {Muller},
  \citenamefont {Neal}, \citenamefont {Mou}, \citenamefont {Thompson},
  \citenamefont {Nair},\ and\ \citenamefont {Schlom}}]{SiO2doping}%
  \BibitemOpen
  \bibfield  {author} {\bibinfo {author} {\bibfnamefont {K.}~\bibnamefont
  {Azizie}}, \bibinfo {author} {\bibfnamefont {F.~V.~E.}\ \bibnamefont
  {Hensling}}, \bibinfo {author} {\bibfnamefont {C.~A.}\ \bibnamefont
  {Gorsak}}, \bibinfo {author} {\bibfnamefont {Y.}~\bibnamefont {Kim}},
  \bibinfo {author} {\bibfnamefont {N.~A.}\ \bibnamefont {Pieczulewski}},
  \bibinfo {author} {\bibfnamefont {D.~M.}\ \bibnamefont {Dryden}}, \bibinfo
  {author} {\bibfnamefont {M.~K.~I.}\ \bibnamefont {Senevirathna}}, \bibinfo
  {author} {\bibfnamefont {S.}~\bibnamefont {Coye}}, \bibinfo {author}
  {\bibfnamefont {S.-L.}\ \bibnamefont {Shang}}, \bibinfo {author}
  {\bibfnamefont {J.}~\bibnamefont {Steele}}, \bibinfo {author} {\bibfnamefont
  {P.}~\bibnamefont {Vogt}}, \bibinfo {author} {\bibfnamefont {N.~A.}\
  \bibnamefont {Parker}}, \bibinfo {author} {\bibfnamefont {Y.~A.}\
  \bibnamefont {Birkhölzer}}, \bibinfo {author} {\bibfnamefont {J.~P.}\
  \bibnamefont {McCandless}}, \bibinfo {author} {\bibfnamefont
  {D.}~\bibnamefont {Jena}}, \bibinfo {author} {\bibfnamefont {H.~G.}\
  \bibnamefont {Xing}}, \bibinfo {author} {\bibfnamefont {Z.-K.}\ \bibnamefont
  {Liu}}, \bibinfo {author} {\bibfnamefont {M.~D.}\ \bibnamefont {Williams}},
  \bibinfo {author} {\bibfnamefont {A.~J.}\ \bibnamefont {Green}}, \bibinfo
  {author} {\bibfnamefont {K.}~\bibnamefont {Chabak}}, \bibinfo {author}
  {\bibfnamefont {D.~A.}\ \bibnamefont {Muller}}, \bibinfo {author}
  {\bibfnamefont {A.~T.}\ \bibnamefont {Neal}}, \bibinfo {author}
  {\bibfnamefont {S.}~\bibnamefont {Mou}}, \bibinfo {author} {\bibfnamefont
  {M.~O.}\ \bibnamefont {Thompson}}, \bibinfo {author} {\bibfnamefont {H.~P.}\
  \bibnamefont {Nair}}, \ and\ \bibinfo {author} {\bibfnamefont {D.~G.}\
  \bibnamefont {Schlom}},\ }\href@noop {} {\bibfield  {journal} {\bibinfo
  {journal} {APL Materials}\ }\textbf {\bibinfo {volume} {11}} (\bibinfo {year}
  {2023})}\BibitemShut {NoStop}%
\bibitem [{\citenamefont {Galazka}\ \emph {et~al.}(2023)\citenamefont
  {Galazka}, \citenamefont {Fiedler}, \citenamefont {Popp}, \citenamefont
  {Ganschow}, \citenamefont {Kwasniewski}, \citenamefont {Seyidov},
  \citenamefont {Pietsch}, \citenamefont {Dittmar}, \citenamefont {Anooz},
  \citenamefont {Irmscher}, \citenamefont {Suendermann}, \citenamefont {Klimm},
  \citenamefont {Chou}, \citenamefont {Rehm}, \citenamefont {Schroeder},\ and\
  \citenamefont {Bickermann}}]{doi:10.1063/5.0131285}%
  \BibitemOpen
  \bibfield  {author} {\bibinfo {author} {\bibfnamefont {Z.}~\bibnamefont
  {Galazka}}, \bibinfo {author} {\bibfnamefont {A.}~\bibnamefont {Fiedler}},
  \bibinfo {author} {\bibfnamefont {A.}~\bibnamefont {Popp}}, \bibinfo {author}
  {\bibfnamefont {S.}~\bibnamefont {Ganschow}}, \bibinfo {author}
  {\bibfnamefont {A.}~\bibnamefont {Kwasniewski}}, \bibinfo {author}
  {\bibfnamefont {P.}~\bibnamefont {Seyidov}}, \bibinfo {author} {\bibfnamefont
  {M.}~\bibnamefont {Pietsch}}, \bibinfo {author} {\bibfnamefont
  {A.}~\bibnamefont {Dittmar}}, \bibinfo {author} {\bibfnamefont {S.~B.}\
  \bibnamefont {Anooz}}, \bibinfo {author} {\bibfnamefont {K.}~\bibnamefont
  {Irmscher}}, \bibinfo {author} {\bibfnamefont {M.}~\bibnamefont
  {Suendermann}}, \bibinfo {author} {\bibfnamefont {D.}~\bibnamefont {Klimm}},
  \bibinfo {author} {\bibfnamefont {T.-S.}\ \bibnamefont {Chou}}, \bibinfo
  {author} {\bibfnamefont {J.}~\bibnamefont {Rehm}}, \bibinfo {author}
  {\bibfnamefont {T.}~\bibnamefont {Schroeder}}, \ and\ \bibinfo {author}
  {\bibfnamefont {M.}~\bibnamefont {Bickermann}},\ }\href@noop {} {\bibfield
  {journal} {\bibinfo  {journal} {Journal of Applied Physics}\ }\textbf
  {\bibinfo {volume} {133}} (\bibinfo {year} {2023})}\BibitemShut {NoStop}%
\bibitem [{\citenamefont {Swallow}\ \emph {et~al.}(2021)\citenamefont
  {Swallow}, \citenamefont {Palgrave}, \citenamefont {Murgatroyd},
  \citenamefont {Regoutz}, \citenamefont {Lorenz}, \citenamefont {Hassa},
  \citenamefont {Grundmann}, \citenamefont {von Wenckstern}, \citenamefont
  {Varley},\ and\ \citenamefont {Veal}}]{doi:10.1021/acsami.0c16021}%
  \BibitemOpen
  \bibfield  {author} {\bibinfo {author} {\bibfnamefont {J.~E.~N.}\
  \bibnamefont {Swallow}}, \bibinfo {author} {\bibfnamefont {R.~G.}\
  \bibnamefont {Palgrave}}, \bibinfo {author} {\bibfnamefont {P.~A.~E.}\
  \bibnamefont {Murgatroyd}}, \bibinfo {author} {\bibfnamefont
  {A.}~\bibnamefont {Regoutz}}, \bibinfo {author} {\bibfnamefont
  {M.}~\bibnamefont {Lorenz}}, \bibinfo {author} {\bibfnamefont
  {A.}~\bibnamefont {Hassa}}, \bibinfo {author} {\bibfnamefont
  {M.}~\bibnamefont {Grundmann}}, \bibinfo {author} {\bibfnamefont
  {H.}~\bibnamefont {von Wenckstern}}, \bibinfo {author} {\bibfnamefont
  {J.~B.}\ \bibnamefont {Varley}}, \ and\ \bibinfo {author} {\bibfnamefont
  {T.~D.}\ \bibnamefont {Veal}},\ }\href@noop {} {\bibfield  {journal}
  {\bibinfo  {journal} {ACS Applied Materials \& Interfaces}\ }\textbf
  {\bibinfo {volume} {13}} (\bibinfo {year} {2021})}\BibitemShut {NoStop}%
\bibitem [{\citenamefont {Mezzadri}\ \emph {et~al.}(2016)\citenamefont
  {Mezzadri}, \citenamefont {Calestani}, \citenamefont {Boschi}, \citenamefont
  {Delmonte}, \citenamefont {Bosi},\ and\ \citenamefont
  {Fornari}}]{mezzadriCrystalStructureFerroelectric2016}%
  \BibitemOpen
  \bibfield  {author} {\bibinfo {author} {\bibfnamefont {F.}~\bibnamefont
  {Mezzadri}}, \bibinfo {author} {\bibfnamefont {G.}~\bibnamefont {Calestani}},
  \bibinfo {author} {\bibfnamefont {F.}~\bibnamefont {Boschi}}, \bibinfo
  {author} {\bibfnamefont {D.}~\bibnamefont {Delmonte}}, \bibinfo {author}
  {\bibfnamefont {M.}~\bibnamefont {Bosi}}, \ and\ \bibinfo {author}
  {\bibfnamefont {R.}~\bibnamefont {Fornari}},\ }\href@noop {} {\bibfield
  {journal} {\bibinfo  {journal} {Inorg. Chem.}\ }\textbf {\bibinfo {volume}
  {55}} (\bibinfo {year} {2016})}\BibitemShut {NoStop}%
\bibitem [{\citenamefont {Maccioni}\ and\ \citenamefont
  {Fiorentini}(2016)}]{maccioniPhaseDiagramPolarization2016a}%
  \BibitemOpen
  \bibfield  {author} {\bibinfo {author} {\bibfnamefont {M.~B.}\ \bibnamefont
  {Maccioni}}\ and\ \bibinfo {author} {\bibfnamefont {V.}~\bibnamefont
  {Fiorentini}},\ }\href@noop {} {\bibfield  {journal} {\bibinfo  {journal}
  {Appl. Phys. Express}\ }\textbf {\bibinfo {volume} {9}} (\bibinfo {year}
  {2016})}\BibitemShut {NoStop}%
\bibitem [{\citenamefont {Ranga}\ \emph {et~al.}(2020)\citenamefont {Ranga},
  \citenamefont {Cho}, \citenamefont {Mishra},\ and\ \citenamefont
  {Krishnamoorthy}}]{rangaHighlyTunablePolarizationengineered2020}%
  \BibitemOpen
  \bibfield  {author} {\bibinfo {author} {\bibfnamefont {P.}~\bibnamefont
  {Ranga}}, \bibinfo {author} {\bibfnamefont {S.~B.}\ \bibnamefont {Cho}},
  \bibinfo {author} {\bibfnamefont {R.}~\bibnamefont {Mishra}}, \ and\ \bibinfo
  {author} {\bibfnamefont {S.}~\bibnamefont {Krishnamoorthy}},\ }\href@noop {}
  {\bibfield  {journal} {\bibinfo  {journal} {Appl. Phys. Express}\ }\textbf
  {\bibinfo {volume} {13}} (\bibinfo {year} {2020})}\BibitemShut {NoStop}%
\bibitem [{\citenamefont {Cho}\ and\ \citenamefont
  {Mishra}(2018)}]{choEpitaxialEngineeringPolar2018}%
  \BibitemOpen
  \bibfield  {author} {\bibinfo {author} {\bibfnamefont {S.~B.}\ \bibnamefont
  {Cho}}\ and\ \bibinfo {author} {\bibfnamefont {R.}~\bibnamefont {Mishra}},\
  }\href@noop {} {\bibfield  {journal} {\bibinfo  {journal} {Appl. Phys.
  Lett.}\ }\textbf {\bibinfo {volume} {112}} (\bibinfo {year}
  {2018})}\BibitemShut {NoStop}%
\bibitem [{\citenamefont {Polyakov}\ \emph {et~al.}(2023)\citenamefont
  {Polyakov}, \citenamefont {Nikolaev}, \citenamefont {Pechnikov},
  \citenamefont {Yakimov}, \citenamefont {Karpov}, \citenamefont {Stepanov},
  \citenamefont {Shchemerov}, \citenamefont {Vasilev}, \citenamefont
  {Chernykh}, \citenamefont {Kuznetsov}, \citenamefont {Lee},\ and\
  \citenamefont {Pearton}}]{POLYAKOV2023168315}%
  \BibitemOpen
  \bibfield  {author} {\bibinfo {author} {\bibfnamefont {A.}~\bibnamefont
  {Polyakov}}, \bibinfo {author} {\bibfnamefont {V.}~\bibnamefont {Nikolaev}},
  \bibinfo {author} {\bibfnamefont {A.}~\bibnamefont {Pechnikov}}, \bibinfo
  {author} {\bibfnamefont {E.}~\bibnamefont {Yakimov}}, \bibinfo {author}
  {\bibfnamefont {S.~Y.}\ \bibnamefont {Karpov}}, \bibinfo {author}
  {\bibfnamefont {S.}~\bibnamefont {Stepanov}}, \bibinfo {author}
  {\bibfnamefont {I.}~\bibnamefont {Shchemerov}}, \bibinfo {author}
  {\bibfnamefont {A.}~\bibnamefont {Vasilev}}, \bibinfo {author} {\bibfnamefont
  {A.}~\bibnamefont {Chernykh}}, \bibinfo {author} {\bibfnamefont
  {A.}~\bibnamefont {Kuznetsov}}, \bibinfo {author} {\bibfnamefont {I.-H.}\
  \bibnamefont {Lee}}, \ and\ \bibinfo {author} {\bibfnamefont
  {S.}~\bibnamefont {Pearton}},\ }\href@noop {} {\bibfield  {journal} {\bibinfo
   {journal} {Journal of Alloys and Compounds}\ }\textbf {\bibinfo {volume}
  {936}} (\bibinfo {year} {2023})}\BibitemShut {NoStop}%
\bibitem [{\citenamefont {Kuang}\ \emph {et~al.}(2021)\citenamefont {Kuang},
  \citenamefont {Chen}, \citenamefont {Ma}, \citenamefont {Du}, \citenamefont
  {Zhang}, \citenamefont {Hao}, \citenamefont {Ren}, \citenamefont {Liu},
  \citenamefont {Zhu}, \citenamefont {Gu}, \citenamefont {Zhang}, \citenamefont
  {Zheng},\ and\ \citenamefont {Ye}}]{kuangBandAlignmentEnhanced2021}%
  \BibitemOpen
  \bibfield  {author} {\bibinfo {author} {\bibfnamefont {Y.}~\bibnamefont
  {Kuang}}, \bibinfo {author} {\bibfnamefont {X.}~\bibnamefont {Chen}},
  \bibinfo {author} {\bibfnamefont {T.}~\bibnamefont {Ma}}, \bibinfo {author}
  {\bibfnamefont {Q.}~\bibnamefont {Du}}, \bibinfo {author} {\bibfnamefont
  {Y.}~\bibnamefont {Zhang}}, \bibinfo {author} {\bibfnamefont
  {J.}~\bibnamefont {Hao}}, \bibinfo {author} {\bibfnamefont {F.-F.}\
  \bibnamefont {Ren}}, \bibinfo {author} {\bibfnamefont {B.}~\bibnamefont
  {Liu}}, \bibinfo {author} {\bibfnamefont {S.}~\bibnamefont {Zhu}}, \bibinfo
  {author} {\bibfnamefont {S.}~\bibnamefont {Gu}}, \bibinfo {author}
  {\bibfnamefont {R.}~\bibnamefont {Zhang}}, \bibinfo {author} {\bibfnamefont
  {Y.}~\bibnamefont {Zheng}}, \ and\ \bibinfo {author} {\bibfnamefont
  {J.}~\bibnamefont {Ye}},\ }\href@noop {} {\bibfield  {journal} {\bibinfo
  {journal} {ACS Appl. Electron. Mater.}\ }\textbf {\bibinfo {volume} {3}}
  (\bibinfo {year} {2021})}\BibitemShut {NoStop}%
\bibitem [{\citenamefont {Mazzolini}\ \emph {et~al.}(2023)\citenamefont
  {Mazzolini}, \citenamefont {Fogarassy}, \citenamefont {Parisini},
  \citenamefont {Mezzadri}, \citenamefont {Diercks}, \citenamefont {Bosi},
  \citenamefont {Seravalli}, \citenamefont {Sacchi}, \citenamefont {Spaggiari},
  \citenamefont {Bersani}, \citenamefont {Bierwagen}, \citenamefont {Janzen},
  \citenamefont {Marggraf}, \citenamefont {Wagner}, \citenamefont {Cora},
  \citenamefont {P{\'e}cz}, \citenamefont {Tahraoui}, \citenamefont {Bosio},
  \citenamefont {Borelli}, \citenamefont {Leone},\ and\ \citenamefont
  {Fornari}}]{mazzoliniSilaneMediatedExpansion2023}%
  \BibitemOpen
  \bibfield  {author} {\bibinfo {author} {\bibfnamefont {P.}~\bibnamefont
  {Mazzolini}}, \bibinfo {author} {\bibfnamefont {Z.}~\bibnamefont
  {Fogarassy}}, \bibinfo {author} {\bibfnamefont {A.}~\bibnamefont {Parisini}},
  \bibinfo {author} {\bibfnamefont {F.}~\bibnamefont {Mezzadri}}, \bibinfo
  {author} {\bibfnamefont {D.}~\bibnamefont {Diercks}}, \bibinfo {author}
  {\bibfnamefont {M.}~\bibnamefont {Bosi}}, \bibinfo {author} {\bibfnamefont
  {L.}~\bibnamefont {Seravalli}}, \bibinfo {author} {\bibfnamefont
  {A.}~\bibnamefont {Sacchi}}, \bibinfo {author} {\bibfnamefont
  {G.}~\bibnamefont {Spaggiari}}, \bibinfo {author} {\bibfnamefont
  {D.}~\bibnamefont {Bersani}}, \bibinfo {author} {\bibfnamefont
  {O.}~\bibnamefont {Bierwagen}}, \bibinfo {author} {\bibfnamefont {B.~M.}\
  \bibnamefont {Janzen}}, \bibinfo {author} {\bibfnamefont {M.~N.}\
  \bibnamefont {Marggraf}}, \bibinfo {author} {\bibfnamefont {M.~R.}\
  \bibnamefont {Wagner}}, \bibinfo {author} {\bibfnamefont {I.}~\bibnamefont
  {Cora}}, \bibinfo {author} {\bibfnamefont {B.}~\bibnamefont {P{\'e}cz}},
  \bibinfo {author} {\bibfnamefont {A.}~\bibnamefont {Tahraoui}}, \bibinfo
  {author} {\bibfnamefont {A.}~\bibnamefont {Bosio}}, \bibinfo {author}
  {\bibfnamefont {C.}~\bibnamefont {Borelli}}, \bibinfo {author} {\bibfnamefont
  {S.}~\bibnamefont {Leone}}, \ and\ \bibinfo {author} {\bibfnamefont
  {R.}~\bibnamefont {Fornari}},\ }\href@noop {} {\bibfield  {journal} {\bibinfo
   {journal} {Adv Funct Materials}\ }\textbf {\bibinfo {volume} {33}} (\bibinfo
  {year} {2023})}\BibitemShut {NoStop}%
\bibitem [{\citenamefont {Oshima}\ \emph {et~al.}(2015)\citenamefont {Oshima},
  \citenamefont {V{\'i}llora}, \citenamefont {Matsushita}, \citenamefont
  {Yamamoto},\ and\ \citenamefont
  {Shimamura}}]{oshimaEpitaxialGrowthPhasepure2015}%
  \BibitemOpen
  \bibfield  {author} {\bibinfo {author} {\bibfnamefont {Y.}~\bibnamefont
  {Oshima}}, \bibinfo {author} {\bibfnamefont {E.~G.}\ \bibnamefont
  {V{\'i}llora}}, \bibinfo {author} {\bibfnamefont {Y.}~\bibnamefont
  {Matsushita}}, \bibinfo {author} {\bibfnamefont {S.}~\bibnamefont
  {Yamamoto}}, \ and\ \bibinfo {author} {\bibfnamefont {K.}~\bibnamefont
  {Shimamura}},\ }\href@noop {} {\bibfield  {journal} {\bibinfo  {journal}
  {Journal of Applied Physics}\ }\textbf {\bibinfo {volume} {118}} (\bibinfo
  {year} {2015})}\BibitemShut {NoStop}%
\bibitem [{\citenamefont {Nishinaka}\ \emph {et~al.}(2020)\citenamefont
  {Nishinaka}, \citenamefont {Ueda}, \citenamefont {Tahara}, \citenamefont
  {Ito}, \citenamefont {Ikenaga}, \citenamefont {Hasuike},\ and\ \citenamefont
  {Yoshimoto}}]{doi:10.1021/acsomega.0c04634}%
  \BibitemOpen
  \bibfield  {author} {\bibinfo {author} {\bibfnamefont {H.}~\bibnamefont
  {Nishinaka}}, \bibinfo {author} {\bibfnamefont {O.}~\bibnamefont {Ueda}},
  \bibinfo {author} {\bibfnamefont {D.}~\bibnamefont {Tahara}}, \bibinfo
  {author} {\bibfnamefont {Y.}~\bibnamefont {Ito}}, \bibinfo {author}
  {\bibfnamefont {N.}~\bibnamefont {Ikenaga}}, \bibinfo {author} {\bibfnamefont
  {N.}~\bibnamefont {Hasuike}}, \ and\ \bibinfo {author} {\bibfnamefont
  {M.}~\bibnamefont {Yoshimoto}},\ }\href@noop {} {\bibfield  {journal}
  {\bibinfo  {journal} {ACS Omega}\ }\textbf {\bibinfo {volume} {5}} (\bibinfo
  {year} {2020})}\BibitemShut {NoStop}%
\bibitem [{\citenamefont {Bosi}\ \emph {et~al.}(2020)\citenamefont {Bosi},
  \citenamefont {Mazzolini}, \citenamefont {Seravalli},\ and\ \citenamefont
  {Fornari}}]{D0TC02743J}%
  \BibitemOpen
  \bibfield  {author} {\bibinfo {author} {\bibfnamefont {M.}~\bibnamefont
  {Bosi}}, \bibinfo {author} {\bibfnamefont {P.}~\bibnamefont {Mazzolini}},
  \bibinfo {author} {\bibfnamefont {L.}~\bibnamefont {Seravalli}}, \ and\
  \bibinfo {author} {\bibfnamefont {R.}~\bibnamefont {Fornari}},\ }\href@noop
  {} {\bibfield  {journal} {\bibinfo  {journal} {J. Mater. Chem. C}\ }\textbf
  {\bibinfo {volume} {8}} (\bibinfo {year} {2020})}\BibitemShut {NoStop}%
\bibitem [{\citenamefont {Knei{\ss}}\ \emph {et~al.}(2019)\citenamefont
  {Knei{\ss}}, \citenamefont {Hassa}, \citenamefont {Splith}, \citenamefont
  {Sturm}, \citenamefont {{von Wenckstern}}, \citenamefont {Schultz},
  \citenamefont {Koch}, \citenamefont {Lorenz},\ and\ \citenamefont
  {Grundmann}}]{kneissTinassistedHeteroepitaxialPLDgrowth2019}%
  \BibitemOpen
  \bibfield  {author} {\bibinfo {author} {\bibfnamefont {M.}~\bibnamefont
  {Knei{\ss}}}, \bibinfo {author} {\bibfnamefont {A.}~\bibnamefont {Hassa}},
  \bibinfo {author} {\bibfnamefont {D.}~\bibnamefont {Splith}}, \bibinfo
  {author} {\bibfnamefont {C.}~\bibnamefont {Sturm}}, \bibinfo {author}
  {\bibfnamefont {H.}~\bibnamefont {{von Wenckstern}}}, \bibinfo {author}
  {\bibfnamefont {T.}~\bibnamefont {Schultz}}, \bibinfo {author} {\bibfnamefont
  {N.}~\bibnamefont {Koch}}, \bibinfo {author} {\bibfnamefont {M.}~\bibnamefont
  {Lorenz}}, \ and\ \bibinfo {author} {\bibfnamefont {M.}~\bibnamefont
  {Grundmann}},\ }\href@noop {} {\bibfield  {journal} {\bibinfo  {journal} {APL
  Materials}\ }\textbf {\bibinfo {volume} {7}} (\bibinfo {year}
  {2019})}\BibitemShut {NoStop}%
\bibitem [{\citenamefont {Nishinaka}, \citenamefont {Tahara},\ and\
  \citenamefont {Yoshimoto}(2016)}]{nishinakaHeteroepitaxialGrowthEGa2016}%
  \BibitemOpen
  \bibfield  {author} {\bibinfo {author} {\bibfnamefont {H.}~\bibnamefont
  {Nishinaka}}, \bibinfo {author} {\bibfnamefont {D.}~\bibnamefont {Tahara}}, \
  and\ \bibinfo {author} {\bibfnamefont {M.}~\bibnamefont {Yoshimoto}},\
  }\href@noop {} {\bibfield  {journal} {\bibinfo  {journal} {Jpn. J. Appl.
  Phys.}\ }\textbf {\bibinfo {volume} {55}} (\bibinfo {year}
  {2016})}\BibitemShut {NoStop}%
\bibitem [{\citenamefont {Vogt}\ \emph {et~al.}(2017)\citenamefont {Vogt},
  \citenamefont {Brandt}, \citenamefont {Riechert}, \citenamefont
  {L{\"a}hnemann},\ and\ \citenamefont
  {Bierwagen}}]{vogtMetalExchangeCatalysisGrowth2017}%
  \BibitemOpen
  \bibfield  {author} {\bibinfo {author} {\bibfnamefont {P.}~\bibnamefont
  {Vogt}}, \bibinfo {author} {\bibfnamefont {O.}~\bibnamefont {Brandt}},
  \bibinfo {author} {\bibfnamefont {H.}~\bibnamefont {Riechert}}, \bibinfo
  {author} {\bibfnamefont {J.}~\bibnamefont {L{\"a}hnemann}}, \ and\ \bibinfo
  {author} {\bibfnamefont {O.}~\bibnamefont {Bierwagen}},\ }\href@noop {}
  {\bibfield  {journal} {\bibinfo  {journal} {Phys. Rev. Lett.}\ }\textbf
  {\bibinfo {volume} {119}} (\bibinfo {year} {2017})}\BibitemShut {NoStop}%
\bibitem [{\citenamefont {Kracht}\ \emph {et~al.}(2017)\citenamefont {Kracht},
  \citenamefont {Karg}, \citenamefont {Sch{\"o}rmann}, \citenamefont
  {Weinhold}, \citenamefont {Zink}, \citenamefont {Michel}, \citenamefont
  {Rohnke}, \citenamefont {Schowalter}, \citenamefont {Gerken}, \citenamefont
  {Rosenauer}, \citenamefont {Klar}, \citenamefont {Janek},\ and\ \citenamefont
  {Eickhoff}}]{krachtTinAssistedSynthesisGa2017}%
  \BibitemOpen
  \bibfield  {author} {\bibinfo {author} {\bibfnamefont {M.}~\bibnamefont
  {Kracht}}, \bibinfo {author} {\bibfnamefont {A.}~\bibnamefont {Karg}},
  \bibinfo {author} {\bibfnamefont {J.}~\bibnamefont {Sch{\"o}rmann}}, \bibinfo
  {author} {\bibfnamefont {M.}~\bibnamefont {Weinhold}}, \bibinfo {author}
  {\bibfnamefont {D.}~\bibnamefont {Zink}}, \bibinfo {author} {\bibfnamefont
  {F.}~\bibnamefont {Michel}}, \bibinfo {author} {\bibfnamefont
  {M.}~\bibnamefont {Rohnke}}, \bibinfo {author} {\bibfnamefont
  {M.}~\bibnamefont {Schowalter}}, \bibinfo {author} {\bibfnamefont
  {B.}~\bibnamefont {Gerken}}, \bibinfo {author} {\bibfnamefont
  {A.}~\bibnamefont {Rosenauer}}, \bibinfo {author} {\bibfnamefont {P.~J.}\
  \bibnamefont {Klar}}, \bibinfo {author} {\bibfnamefont {J.}~\bibnamefont
  {Janek}}, \ and\ \bibinfo {author} {\bibfnamefont {M.}~\bibnamefont
  {Eickhoff}},\ }\href@noop {} {\bibfield  {journal} {\bibinfo  {journal}
  {Phys. Rev. Applied}\ }\textbf {\bibinfo {volume} {8}} (\bibinfo {year}
  {2017})}\BibitemShut {NoStop}%
\bibitem [{\citenamefont {Boschi}\ \emph {et~al.}(2016)\citenamefont {Boschi},
  \citenamefont {Bosi}, \citenamefont {Berzina}, \citenamefont {Buffagni},
  \citenamefont {Ferrari},\ and\ \citenamefont
  {Fornari}}]{boschi_hetero-epitaxy_2016}%
  \BibitemOpen
  \bibfield  {author} {\bibinfo {author} {\bibfnamefont {F.}~\bibnamefont
  {Boschi}}, \bibinfo {author} {\bibfnamefont {M.}~\bibnamefont {Bosi}},
  \bibinfo {author} {\bibfnamefont {T.}~\bibnamefont {Berzina}}, \bibinfo
  {author} {\bibfnamefont {E.}~\bibnamefont {Buffagni}}, \bibinfo {author}
  {\bibfnamefont {C.}~\bibnamefont {Ferrari}}, \ and\ \bibinfo {author}
  {\bibfnamefont {R.}~\bibnamefont {Fornari}},\ }\href@noop {} {\bibfield
  {journal} {\bibinfo  {journal} {Journal of Crystal Growth}\ }\textbf
  {\bibinfo {volume} {443}} (\bibinfo {year} {2016})}\BibitemShut {NoStop}%
\bibitem [{\citenamefont {Mazzolini}\ \emph
  {et~al.}(2020{\natexlab{a}})\citenamefont {Mazzolini}, \citenamefont
  {Falkenstein}, \citenamefont {Wouters}, \citenamefont {Schewski},
  \citenamefont {Markurt}, \citenamefont {Galazka}, \citenamefont {Martin},
  \citenamefont {Albrecht},\ and\ \citenamefont
  {Bierwagen}}]{mazzoliniSubstrateorientationDependenceVGa2020}%
  \BibitemOpen
  \bibfield  {author} {\bibinfo {author} {\bibfnamefont {P.}~\bibnamefont
  {Mazzolini}}, \bibinfo {author} {\bibfnamefont {A.}~\bibnamefont
  {Falkenstein}}, \bibinfo {author} {\bibfnamefont {C.}~\bibnamefont
  {Wouters}}, \bibinfo {author} {\bibfnamefont {R.}~\bibnamefont {Schewski}},
  \bibinfo {author} {\bibfnamefont {T.}~\bibnamefont {Markurt}}, \bibinfo
  {author} {\bibfnamefont {Z.}~\bibnamefont {Galazka}}, \bibinfo {author}
  {\bibfnamefont {M.}~\bibnamefont {Martin}}, \bibinfo {author} {\bibfnamefont
  {M.}~\bibnamefont {Albrecht}}, \ and\ \bibinfo {author} {\bibfnamefont
  {O.}~\bibnamefont {Bierwagen}},\ }\href@noop {} {\bibfield  {journal}
  {\bibinfo  {journal} {APL Materials}\ }\textbf {\bibinfo {volume} {8}}
  (\bibinfo {year} {2020}{\natexlab{a}})}\BibitemShut {NoStop}%
\bibitem [{\citenamefont {Mazzolini}\ \emph
  {et~al.}(2020{\natexlab{b}})\citenamefont {Mazzolini}, \citenamefont
  {Falkenstein}, \citenamefont {Galazka}, \citenamefont {Martin},\ and\
  \citenamefont {Bierwagen}}]{mazzolini_offcut-related_2020}%
  \BibitemOpen
  \bibfield  {author} {\bibinfo {author} {\bibfnamefont {P.}~\bibnamefont
  {Mazzolini}}, \bibinfo {author} {\bibfnamefont {A.}~\bibnamefont
  {Falkenstein}}, \bibinfo {author} {\bibfnamefont {Z.}~\bibnamefont
  {Galazka}}, \bibinfo {author} {\bibfnamefont {M.}~\bibnamefont {Martin}}, \
  and\ \bibinfo {author} {\bibfnamefont {O.}~\bibnamefont {Bierwagen}},\
  }\href@noop {} {\bibfield  {journal} {\bibinfo  {journal} {Appl. Phys.
  Lett.}\ }\textbf {\bibinfo {volume} {117}} (\bibinfo {year}
  {2020}{\natexlab{b}})}\BibitemShut {NoStop}%
\bibitem [{\citenamefont {Karg}\ \emph {et~al.}(2022)\citenamefont {Karg},
  \citenamefont {Kracht}, \citenamefont {Vogt}, \citenamefont {Messow},
  \citenamefont {Braud}, \citenamefont {Sch{\"o}rmann}, \citenamefont {Rohnke},
  \citenamefont {Janek}, \citenamefont {Falta},\ and\ \citenamefont
  {Eickhoff}}]{kargEnhancedEpitaxialGrowth2022a}%
  \BibitemOpen
  \bibfield  {author} {\bibinfo {author} {\bibfnamefont {A.}~\bibnamefont
  {Karg}}, \bibinfo {author} {\bibfnamefont {M.}~\bibnamefont {Kracht}},
  \bibinfo {author} {\bibfnamefont {P.}~\bibnamefont {Vogt}}, \bibinfo {author}
  {\bibfnamefont {A.}~\bibnamefont {Messow}}, \bibinfo {author} {\bibfnamefont
  {N.}~\bibnamefont {Braud}}, \bibinfo {author} {\bibfnamefont
  {J.}~\bibnamefont {Sch{\"o}rmann}}, \bibinfo {author} {\bibfnamefont
  {M.}~\bibnamefont {Rohnke}}, \bibinfo {author} {\bibfnamefont
  {J.}~\bibnamefont {Janek}}, \bibinfo {author} {\bibfnamefont
  {J.}~\bibnamefont {Falta}}, \ and\ \bibinfo {author} {\bibfnamefont
  {M.}~\bibnamefont {Eickhoff}},\ }\href@noop {} {\bibfield  {journal}
  {\bibinfo  {journal} {Journal of Applied Physics}\ }\textbf {\bibinfo
  {volume} {132}} (\bibinfo {year} {2022})}\BibitemShut {NoStop}%
\bibitem [{\citenamefont {McCandless}\ \emph {et~al.}(2023)\citenamefont
  {McCandless}, \citenamefont {Rowe}, \citenamefont {Pieczulewski},
  \citenamefont {Protasenko}, \citenamefont {{Alonso-Orts}}, \citenamefont
  {Williams}, \citenamefont {Eickhoff}, \citenamefont {Xing}, \citenamefont
  {Muller}, \citenamefont {Jena},\ and\ \citenamefont
  {Vogt}}]{mccandlessGrowthAGaAAl2023}%
  \BibitemOpen
  \bibfield  {author} {\bibinfo {author} {\bibfnamefont {J.~P.}\ \bibnamefont
  {McCandless}}, \bibinfo {author} {\bibfnamefont {D.}~\bibnamefont {Rowe}},
  \bibinfo {author} {\bibfnamefont {N.}~\bibnamefont {Pieczulewski}}, \bibinfo
  {author} {\bibfnamefont {V.}~\bibnamefont {Protasenko}}, \bibinfo {author}
  {\bibfnamefont {M.}~\bibnamefont {{Alonso-Orts}}}, \bibinfo {author}
  {\bibfnamefont {M.~S.}\ \bibnamefont {Williams}}, \bibinfo {author}
  {\bibfnamefont {M.}~\bibnamefont {Eickhoff}}, \bibinfo {author}
  {\bibfnamefont {H.~G.}\ \bibnamefont {Xing}}, \bibinfo {author}
  {\bibfnamefont {D.~A.}\ \bibnamefont {Muller}}, \bibinfo {author}
  {\bibfnamefont {D.}~\bibnamefont {Jena}}, \ and\ \bibinfo {author}
  {\bibfnamefont {P.}~\bibnamefont {Vogt}},\ }\href@noop {} {\bibfield
  {journal} {\bibinfo  {journal} {Jpn. J. Appl. Phys.}\ }\textbf {\bibinfo
  {volume} {62}} (\bibinfo {year} {2023})}\BibitemShut {NoStop}%
\bibitem [{\citenamefont {Mauze}\ \emph
  {et~al.}(2020{\natexlab{b}})\citenamefont {Mauze}, \citenamefont {Zhang},
  \citenamefont {Itoh}, \citenamefont {Wu},\ and\ \citenamefont
  {Speck}}]{mauzeMetalOxideCatalyzed2020}%
  \BibitemOpen
  \bibfield  {author} {\bibinfo {author} {\bibfnamefont {A.}~\bibnamefont
  {Mauze}}, \bibinfo {author} {\bibfnamefont {Y.}~\bibnamefont {Zhang}},
  \bibinfo {author} {\bibfnamefont {T.}~\bibnamefont {Itoh}}, \bibinfo {author}
  {\bibfnamefont {F.}~\bibnamefont {Wu}}, \ and\ \bibinfo {author}
  {\bibfnamefont {J.~S.}\ \bibnamefont {Speck}},\ }\href@noop {} {\bibfield
  {journal} {\bibinfo  {journal} {APL Materials}\ }\textbf {\bibinfo {volume}
  {8}} (\bibinfo {year} {2020}{\natexlab{b}})}\BibitemShut {NoStop}%
\bibitem [{\citenamefont {Karg}\ \emph {et~al.}(2023)\citenamefont {Karg},
  \citenamefont {Hinz}, \citenamefont {Figge}, \citenamefont {Schowalter},
  \citenamefont {Vogt}, \citenamefont {Rosenauer},\ and\ \citenamefont
  {Eickhoff}}]{IndiumSurfKarg}%
  \BibitemOpen
  \bibfield  {author} {\bibinfo {author} {\bibfnamefont {A.}~\bibnamefont
  {Karg}}, \bibinfo {author} {\bibfnamefont {A.}~\bibnamefont {Hinz}}, \bibinfo
  {author} {\bibfnamefont {S.}~\bibnamefont {Figge}}, \bibinfo {author}
  {\bibfnamefont {M.}~\bibnamefont {Schowalter}}, \bibinfo {author}
  {\bibfnamefont {P.}~\bibnamefont {Vogt}}, \bibinfo {author} {\bibfnamefont
  {A.}~\bibnamefont {Rosenauer}}, \ and\ \bibinfo {author} {\bibfnamefont
  {M.}~\bibnamefont {Eickhoff}},\ }\href {\doibase 10.1063/5.0167736}
  {\bibfield  {journal} {\bibinfo  {journal} {APL Materials}\ }\textbf
  {\bibinfo {volume} {11}},\ \bibinfo {pages} {091114} (\bibinfo {year}
  {2023})}\BibitemShut {NoStop}%
\bibitem [{\citenamefont {Hassa}\ \emph {et~al.}(2020)\citenamefont {Hassa},
  \citenamefont {Wouters}, \citenamefont {Kneiß}, \citenamefont {Splith},
  \citenamefont {Sturm}, \citenamefont {Von~Wenckstern}, \citenamefont
  {Albrecht}, \citenamefont {Lorenz},\ and\ \citenamefont
  {Grundmann}}]{hassaControlPhaseFormation2020}%
  \BibitemOpen
  \bibfield  {author} {\bibinfo {author} {\bibfnamefont {A.}~\bibnamefont
  {Hassa}}, \bibinfo {author} {\bibfnamefont {C.}~\bibnamefont {Wouters}},
  \bibinfo {author} {\bibfnamefont {M.}~\bibnamefont {Kneiß}}, \bibinfo
  {author} {\bibfnamefont {D.}~\bibnamefont {Splith}}, \bibinfo {author}
  {\bibfnamefont {C.}~\bibnamefont {Sturm}}, \bibinfo {author} {\bibfnamefont
  {H.}~\bibnamefont {Von~Wenckstern}}, \bibinfo {author} {\bibfnamefont
  {M.}~\bibnamefont {Albrecht}}, \bibinfo {author} {\bibfnamefont
  {M.}~\bibnamefont {Lorenz}}, \ and\ \bibinfo {author} {\bibfnamefont
  {M.}~\bibnamefont {Grundmann}},\ }\href@noop {} {\bibfield  {journal}
  {\bibinfo  {journal} {J. Phys. D: Appl. Phys.}\ }\textbf {\bibinfo {volume}
  {53}} (\bibinfo {year} {2020})}\BibitemShut {NoStop}%
\bibitem [{\citenamefont {Mazzolini}\ \emph {et~al.}(2019)\citenamefont
  {Mazzolini}, \citenamefont {Vogt}, \citenamefont {Schewski}, \citenamefont
  {Wouters}, \citenamefont {Albrecht},\ and\ \citenamefont
  {Bierwagen}}]{mazzoliniFacetingMetalexchangeCatalysis2019}%
  \BibitemOpen
  \bibfield  {author} {\bibinfo {author} {\bibfnamefont {P.}~\bibnamefont
  {Mazzolini}}, \bibinfo {author} {\bibfnamefont {P.}~\bibnamefont {Vogt}},
  \bibinfo {author} {\bibfnamefont {R.}~\bibnamefont {Schewski}}, \bibinfo
  {author} {\bibfnamefont {C.}~\bibnamefont {Wouters}}, \bibinfo {author}
  {\bibfnamefont {M.}~\bibnamefont {Albrecht}}, \ and\ \bibinfo {author}
  {\bibfnamefont {O.}~\bibnamefont {Bierwagen}},\ }\href@noop {} {\bibfield
  {journal} {\bibinfo  {journal} {APL Materials}\ }\textbf {\bibinfo {volume}
  {7}} (\bibinfo {year} {2019})}\BibitemShut {NoStop}%
\bibitem [{\citenamefont {Fornari}\ \emph {et~al.}(2017)\citenamefont
  {Fornari}, \citenamefont {Pavesi}, \citenamefont {Montedoro}, \citenamefont
  {Klimm}, \citenamefont {Mezzadri}, \citenamefont {Cora}, \citenamefont
  {P{\'e}cz}, \citenamefont {Boschi}, \citenamefont {Parisini}, \citenamefont
  {Baraldi}, \citenamefont {Ferrari}, \citenamefont {Gombia},\ and\
  \citenamefont {Bosi}}]{fornariThermalStabilityEGa2O32017}%
  \BibitemOpen
  \bibfield  {author} {\bibinfo {author} {\bibfnamefont {R.}~\bibnamefont
  {Fornari}}, \bibinfo {author} {\bibfnamefont {M.}~\bibnamefont {Pavesi}},
  \bibinfo {author} {\bibfnamefont {V.}~\bibnamefont {Montedoro}}, \bibinfo
  {author} {\bibfnamefont {D.}~\bibnamefont {Klimm}}, \bibinfo {author}
  {\bibfnamefont {F.}~\bibnamefont {Mezzadri}}, \bibinfo {author}
  {\bibfnamefont {I.}~\bibnamefont {Cora}}, \bibinfo {author} {\bibfnamefont
  {B.}~\bibnamefont {P{\'e}cz}}, \bibinfo {author} {\bibfnamefont
  {F.}~\bibnamefont {Boschi}}, \bibinfo {author} {\bibfnamefont
  {A.}~\bibnamefont {Parisini}}, \bibinfo {author} {\bibfnamefont
  {A.}~\bibnamefont {Baraldi}}, \bibinfo {author} {\bibfnamefont
  {C.}~\bibnamefont {Ferrari}}, \bibinfo {author} {\bibfnamefont
  {E.}~\bibnamefont {Gombia}}, \ and\ \bibinfo {author} {\bibfnamefont
  {M.}~\bibnamefont {Bosi}},\ }\href@noop {} {\bibfield  {journal} {\bibinfo
  {journal} {Acta Materialia}\ }\textbf {\bibinfo {volume} {140}} (\bibinfo
  {year} {2017})}\BibitemShut {NoStop}%
\bibitem [{\citenamefont {Momma}\ and\ \citenamefont
  {Izumi}(2011)}]{mommaVESTAThreedimensionalVisualization2011}%
  \BibitemOpen
  \bibfield  {author} {\bibinfo {author} {\bibfnamefont {K.}~\bibnamefont
  {Momma}}\ and\ \bibinfo {author} {\bibfnamefont {F.}~\bibnamefont {Izumi}},\
  }\href@noop {} {\bibfield  {journal} {\bibinfo  {journal} {J Appl
  Crystallogr}\ }\textbf {\bibinfo {volume} {44}},\ \bibinfo {pages}
  {1272--1276} (\bibinfo {year} {2011})}\BibitemShut {NoStop}%
\bibitem [{\citenamefont {Reis}\ \emph {et~al.}(2023)\citenamefont {Reis},
  \citenamefont {Hanke}, \citenamefont {Bierwagen}, \citenamefont {Trampert},
  \citenamefont {Mazzolini},\ and\ \citenamefont
  {Welter}}]{reisDisorderOrderTransition2023}%
  \BibitemOpen
  \bibfield  {author} {\bibinfo {author} {\bibfnamefont {A.}~\bibnamefont
  {Reis}}, \bibinfo {author} {\bibfnamefont {M.}~\bibnamefont {Hanke}},
  \bibinfo {author} {\bibfnamefont {O.}~\bibnamefont {Bierwagen}}, \bibinfo
  {author} {\bibfnamefont {A.}~\bibnamefont {Trampert}}, \bibinfo {author}
  {\bibfnamefont {P.}~\bibnamefont {Mazzolini}}, \ and\ \bibinfo {author}
  {\bibfnamefont {E.}~\bibnamefont {Welter}},\ }\href@noop {} {\ \textbf
  {\bibinfo {volume} {260}} (\bibinfo {year} {2023})}\BibitemShut {NoStop}%
\bibitem [{\citenamefont {Ratcliff}\ \emph {et~al.}(2022)\citenamefont
  {Ratcliff}, \citenamefont {Oshima}, \citenamefont {Nippert}, \citenamefont
  {Janzen}, \citenamefont {Kluth}, \citenamefont {Goldhahn}, \citenamefont
  {Feneberg}, \citenamefont {Mazzolini}, \citenamefont {Bierwagen},
  \citenamefont {Wouters}, \citenamefont {Nofal}, \citenamefont {Albrecht},
  \citenamefont {Swallow}, \citenamefont {Jones}, \citenamefont {Thakur},
  \citenamefont {Lee}, \citenamefont {Kalha}, \citenamefont {Schlueter},
  \citenamefont {Veal}, \citenamefont {Varley}, \citenamefont {Wagner},\ and\
  \citenamefont {Regoutz}}]{GammaDisorder}%
  \BibitemOpen
  \bibfield  {author} {\bibinfo {author} {\bibfnamefont {L.~E.}\ \bibnamefont
  {Ratcliff}}, \bibinfo {author} {\bibfnamefont {T.}~\bibnamefont {Oshima}},
  \bibinfo {author} {\bibfnamefont {F.}~\bibnamefont {Nippert}}, \bibinfo
  {author} {\bibfnamefont {B.~M.}\ \bibnamefont {Janzen}}, \bibinfo {author}
  {\bibfnamefont {E.}~\bibnamefont {Kluth}}, \bibinfo {author} {\bibfnamefont
  {R.}~\bibnamefont {Goldhahn}}, \bibinfo {author} {\bibfnamefont
  {M.}~\bibnamefont {Feneberg}}, \bibinfo {author} {\bibfnamefont
  {P.}~\bibnamefont {Mazzolini}}, \bibinfo {author} {\bibfnamefont
  {O.}~\bibnamefont {Bierwagen}}, \bibinfo {author} {\bibfnamefont
  {C.}~\bibnamefont {Wouters}}, \bibinfo {author} {\bibfnamefont
  {M.}~\bibnamefont {Nofal}}, \bibinfo {author} {\bibfnamefont
  {M.}~\bibnamefont {Albrecht}}, \bibinfo {author} {\bibfnamefont {J.~E.~N.}\
  \bibnamefont {Swallow}}, \bibinfo {author} {\bibfnamefont {L.~A.~H.}\
  \bibnamefont {Jones}}, \bibinfo {author} {\bibfnamefont {P.~K.}\ \bibnamefont
  {Thakur}}, \bibinfo {author} {\bibfnamefont {T.-L.}\ \bibnamefont {Lee}},
  \bibinfo {author} {\bibfnamefont {C.}~\bibnamefont {Kalha}}, \bibinfo
  {author} {\bibfnamefont {C.}~\bibnamefont {Schlueter}}, \bibinfo {author}
  {\bibfnamefont {T.~D.}\ \bibnamefont {Veal}}, \bibinfo {author}
  {\bibfnamefont {J.~B.}\ \bibnamefont {Varley}}, \bibinfo {author}
  {\bibfnamefont {M.~R.}\ \bibnamefont {Wagner}}, \ and\ \bibinfo {author}
  {\bibfnamefont {A.}~\bibnamefont {Regoutz}},\ }\href@noop {} {\bibfield
  {journal} {\bibinfo  {journal} {Advanced Materials}\ }\textbf {\bibinfo
  {volume} {34}},\ \bibinfo {pages} {2204217} (\bibinfo {year}
  {2022})}\BibitemShut {NoStop}%
\bibitem [{\citenamefont {Cora}\ \emph {et~al.}(2017)\citenamefont {Cora},
  \citenamefont {Mezzadri}, \citenamefont {Boschi}, \citenamefont {Bosi},
  \citenamefont {{\v C}aplovi{\v c}ov{\'a}}, \citenamefont {Calestani},
  \citenamefont {D{\'o}dony}, \citenamefont {P{\'e}cz},\ and\ \citenamefont
  {Fornari}}]{coraRealStructureEGa2017}%
  \BibitemOpen
  \bibfield  {author} {\bibinfo {author} {\bibfnamefont {I.}~\bibnamefont
  {Cora}}, \bibinfo {author} {\bibfnamefont {F.}~\bibnamefont {Mezzadri}},
  \bibinfo {author} {\bibfnamefont {F.}~\bibnamefont {Boschi}}, \bibinfo
  {author} {\bibfnamefont {M.}~\bibnamefont {Bosi}}, \bibinfo {author}
  {\bibfnamefont {M.}~\bibnamefont {{\v C}aplovi{\v c}ov{\'a}}}, \bibinfo
  {author} {\bibfnamefont {G.}~\bibnamefont {Calestani}}, \bibinfo {author}
  {\bibfnamefont {I.}~\bibnamefont {D{\'o}dony}}, \bibinfo {author}
  {\bibfnamefont {B.}~\bibnamefont {P{\'e}cz}}, \ and\ \bibinfo {author}
  {\bibfnamefont {R.}~\bibnamefont {Fornari}},\ }\href@noop {} {\bibfield
  {journal} {\bibinfo  {journal} {CrystEngComm}\ }\textbf {\bibinfo {volume}
  {19}} (\bibinfo {year} {2017})}\BibitemShut {NoStop}%
\bibitem [{\citenamefont {Nishinaka}\ \emph {et~al.}(2018)\citenamefont
  {Nishinaka}, \citenamefont {Miyauchi}, \citenamefont {Tahara}, \citenamefont
  {Morimoto},\ and\ \citenamefont
  {Yoshimoto}}]{nishinakaIncorporationIndiumEgallium2018}%
  \BibitemOpen
  \bibfield  {author} {\bibinfo {author} {\bibfnamefont {H.}~\bibnamefont
  {Nishinaka}}, \bibinfo {author} {\bibfnamefont {N.}~\bibnamefont {Miyauchi}},
  \bibinfo {author} {\bibfnamefont {D.}~\bibnamefont {Tahara}}, \bibinfo
  {author} {\bibfnamefont {S.}~\bibnamefont {Morimoto}}, \ and\ \bibinfo
  {author} {\bibfnamefont {M.}~\bibnamefont {Yoshimoto}},\ }\href@noop {}
  {\bibfield  {journal} {\bibinfo  {journal} {CrystEngComm}\ }\textbf {\bibinfo
  {volume} {20}} (\bibinfo {year} {2018})}\BibitemShut {NoStop}%
\bibitem [{\citenamefont {Kneiß}\ \emph {et~al.}(2019)\citenamefont {Kneiß},
  \citenamefont {Hassa}, \citenamefont {Splith}, \citenamefont {Sturm},
  \citenamefont {von Wenckstern}, \citenamefont {Lorenz},\ and\ \citenamefont
  {Grundmann}}]{PLDupto28}%
  \BibitemOpen
  \bibfield  {author} {\bibinfo {author} {\bibfnamefont {M.}~\bibnamefont
  {Kneiß}}, \bibinfo {author} {\bibfnamefont {A.}~\bibnamefont {Hassa}},
  \bibinfo {author} {\bibfnamefont {D.}~\bibnamefont {Splith}}, \bibinfo
  {author} {\bibfnamefont {C.}~\bibnamefont {Sturm}}, \bibinfo {author}
  {\bibfnamefont {H.}~\bibnamefont {von Wenckstern}}, \bibinfo {author}
  {\bibfnamefont {M.}~\bibnamefont {Lorenz}}, \ and\ \bibinfo {author}
  {\bibfnamefont {M.}~\bibnamefont {Grundmann}},\ }\href {\doibase
  10.1063/1.5120578} {\bibfield  {journal} {\bibinfo  {journal} {APL
  Materials}\ }\textbf {\bibinfo {volume} {7}},\ \bibinfo {pages} {101102}
  (\bibinfo {year} {2019})}\BibitemShut {NoStop}%
\bibitem [{\citenamefont {Nakagomi}\ and\ \citenamefont
  {Kokubun}(2013)}]{https://doi.org/10.1002/pssa.201329040}%
  \BibitemOpen
  \bibfield  {author} {\bibinfo {author} {\bibfnamefont {S.}~\bibnamefont
  {Nakagomi}}\ and\ \bibinfo {author} {\bibfnamefont {Y.}~\bibnamefont
  {Kokubun}},\ }\href {\doibase https://doi.org/10.1002/pssa.201329040}
  {\bibfield  {journal} {\bibinfo  {journal} {physica status solidi (a)}\
  }\textbf {\bibinfo {volume} {210}},\ \bibinfo {pages} {1738--1744} (\bibinfo
  {year} {2013})}\BibitemShut {NoStop}%
\bibitem [{\citenamefont {Saadatkia}\ \emph {et~al.}(2020)\citenamefont
  {Saadatkia}, \citenamefont {Agarwal}, \citenamefont {Hernandez},
  \citenamefont {Reed}, \citenamefont {Brackenbury}, \citenamefont {Codding},
  \citenamefont {Liedke}, \citenamefont {Butterling}, \citenamefont {Wagner},\
  and\ \citenamefont {Selim}}]{PhysRevMaterials.4.104602}%
  \BibitemOpen
  \bibfield  {author} {\bibinfo {author} {\bibfnamefont {P.}~\bibnamefont
  {Saadatkia}}, \bibinfo {author} {\bibfnamefont {S.}~\bibnamefont {Agarwal}},
  \bibinfo {author} {\bibfnamefont {A.}~\bibnamefont {Hernandez}}, \bibinfo
  {author} {\bibfnamefont {E.}~\bibnamefont {Reed}}, \bibinfo {author}
  {\bibfnamefont {I.~D.}\ \bibnamefont {Brackenbury}}, \bibinfo {author}
  {\bibfnamefont {C.~L.}\ \bibnamefont {Codding}}, \bibinfo {author}
  {\bibfnamefont {M.~O.}\ \bibnamefont {Liedke}}, \bibinfo {author}
  {\bibfnamefont {M.}~\bibnamefont {Butterling}}, \bibinfo {author}
  {\bibfnamefont {A.}~\bibnamefont {Wagner}}, \ and\ \bibinfo {author}
  {\bibfnamefont {F.~A.}\ \bibnamefont {Selim}},\ }\href {\doibase
  10.1103/PhysRevMaterials.4.104602} {\bibfield  {journal} {\bibinfo  {journal}
  {Phys. Rev. Mater.}\ }\textbf {\bibinfo {volume} {4}},\ \bibinfo {pages}
  {104602} (\bibinfo {year} {2020})}\BibitemShut {NoStop}%
\bibitem [{\citenamefont {Rafique}\ \emph {et~al.}(2018)\citenamefont
  {Rafique}, \citenamefont {Han}, \citenamefont {Neal}, \citenamefont {Mou},
  \citenamefont {Boeckl},\ and\ \citenamefont {Zhao}}]{rafique_towards_2018}%
  \BibitemOpen
  \bibfield  {author} {\bibinfo {author} {\bibfnamefont {S.}~\bibnamefont
  {Rafique}}, \bibinfo {author} {\bibfnamefont {L.}~\bibnamefont {Han}},
  \bibinfo {author} {\bibfnamefont {A.~T.}\ \bibnamefont {Neal}}, \bibinfo
  {author} {\bibfnamefont {S.}~\bibnamefont {Mou}}, \bibinfo {author}
  {\bibfnamefont {J.}~\bibnamefont {Boeckl}}, \ and\ \bibinfo {author}
  {\bibfnamefont {H.}~\bibnamefont {Zhao}},\ }\href@noop {} {\bibfield
  {journal} {\bibinfo  {journal} {Phys. Status Solidi A}\ }\textbf {\bibinfo
  {volume} {215}} (\bibinfo {year} {2018})}\BibitemShut {NoStop}%
\end{thebibliography}%

%%%%%%%%%%%%%%%% SUPPLEMENTARY

\setcounter{figure}{0}
\pagebreak
\clearpage

\onecolumngrid
\begin{center}
  \textbf{\large Supplementary Material\\
  Phase-selective growth of $\kappa$- vs $\beta$-Ga$_2$O$_3$ and (In$_x$Ga$_{1-x}$)$_2$O$_3$ by In-mediated metal exchange catalysis in plasma-assisted molecular beam epitaxy}\\[.2cm]
\end{center}

\begin{figure*} [h!]
\includegraphics[width=0.7\linewidth]{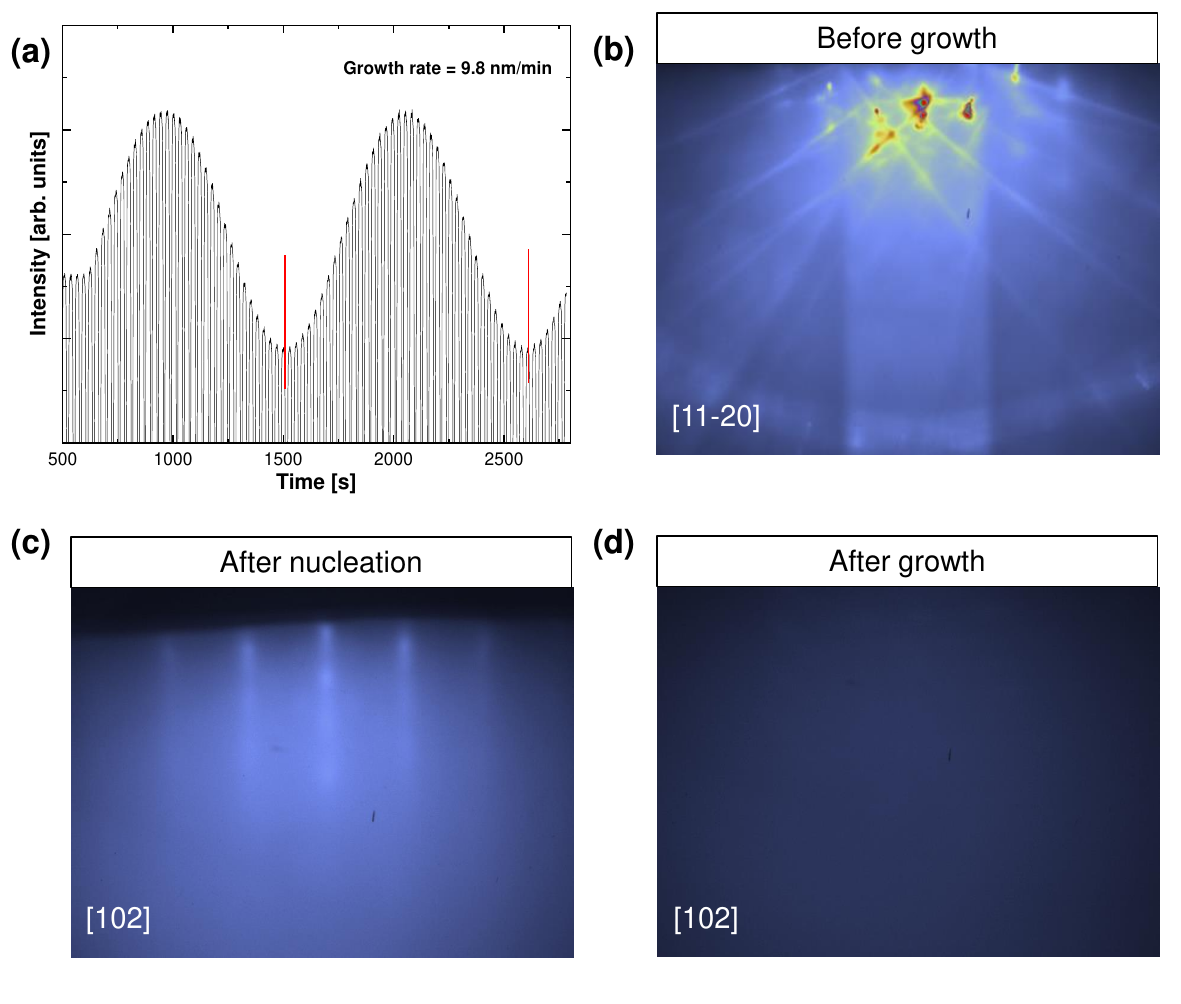}
\caption{\label{Supp1} (a) Laser reflectometry showing the growth rate of an epilayer object of this study (sample A). RHEED pattern of the (b) $\alpha$-Al$_2$O$_3$ substrate, (c) after the growth of the $\beta$-Ga$_2$O$_3$ ($\bar{2}$01) nucleation layer and (d) after the MEXCAT growth of an amorphous layer (sample A).}
\end{figure*}

\begin{figure*}[h!]
\includegraphics[width=0.85\linewidth]{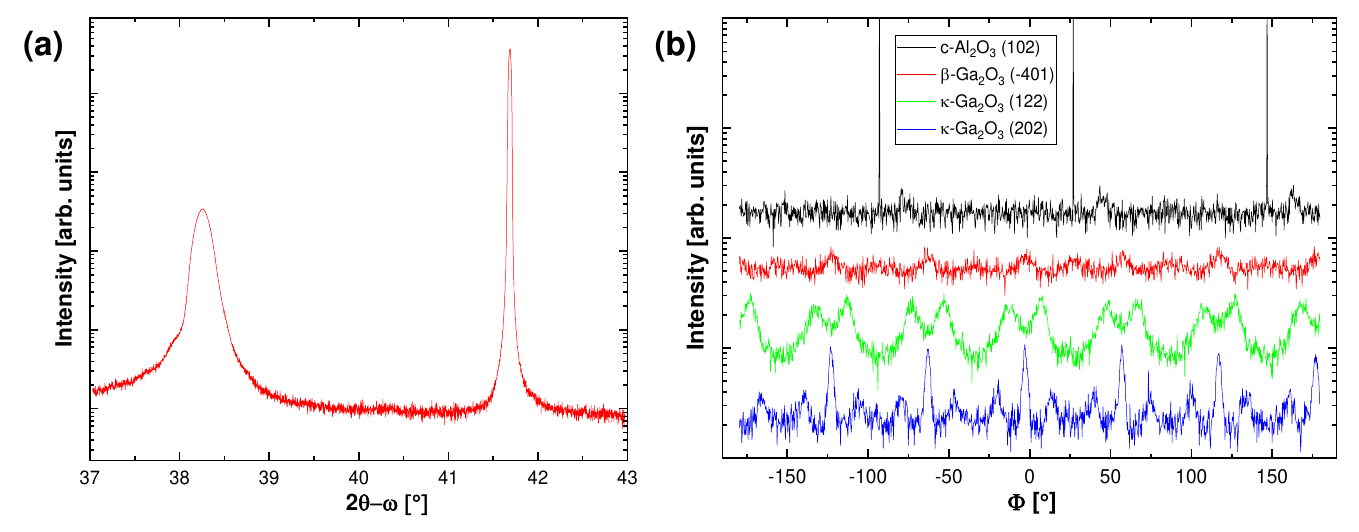}
\caption{\label{Supp2} (a) XRD 2$\theta$-$\omega$ and (b) $\phi$ scan referred to a (In$_x$Ga$_{1-x}$)$_2$O$_3$ (sample B).}
\end{figure*}

\begin{figure*}[h!]
\includegraphics[width=0.8\linewidth]{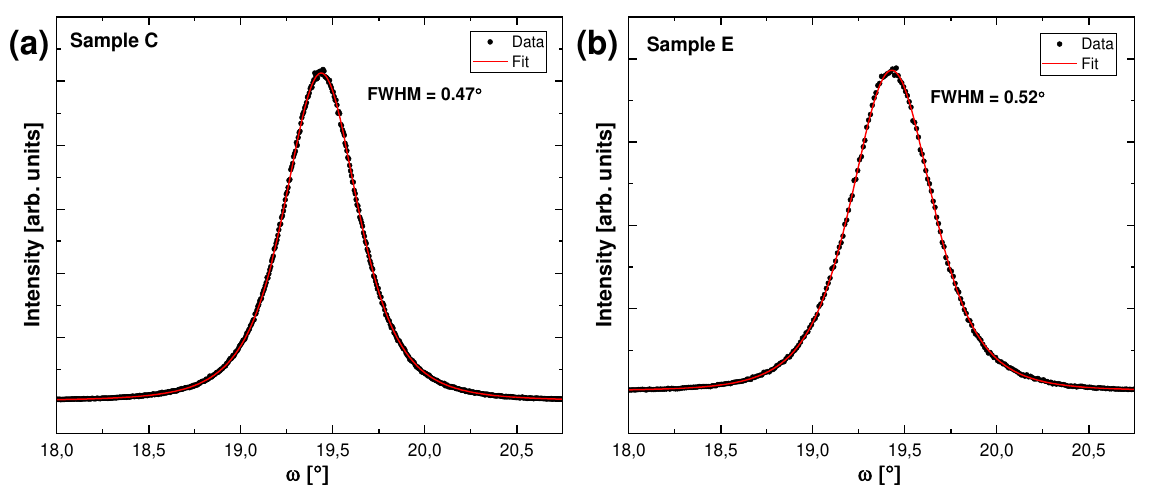}
\caption{\label{Supp3} XRD rocking-curves and respective FWHM of the $\kappa$-Ga$_2$O$_3$ 004 reflex for sample C (a) and sample E (b).}
\end{figure*}

\begin{figure*}[h!]
\includegraphics[width=0.6\linewidth]{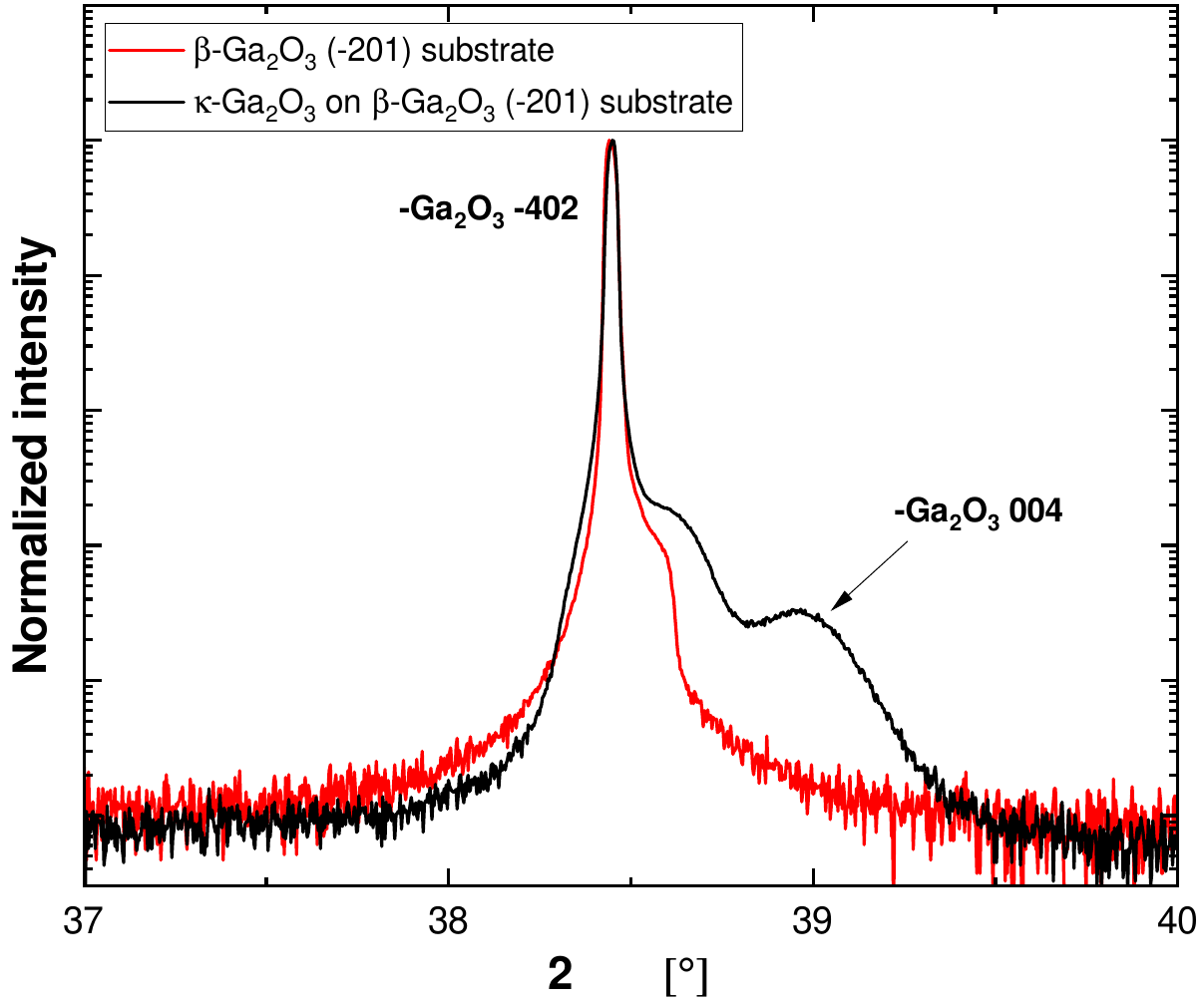}
\caption{\label{Supp4} XRD 2$\theta$-$\omega$ scan of a bulk $\beta$-Ga$_2$O$_3$ ($\bar{2}$01) substrate (red) and of the $\kappa$-Ga$_2$O$_3$ (001) grown on top of a $\beta$-Ga$_2$O$_3$ ($\bar{2}$01) substrate (black). Both the scan were performed for the same in-plane orientation aligning using the -401 reflex of $\beta$-Ga$_2$O$_3$ ($\bar{2}$01)}
\end{figure*}

\begin{figure*}[h!]
\includegraphics[width=0.9\linewidth]{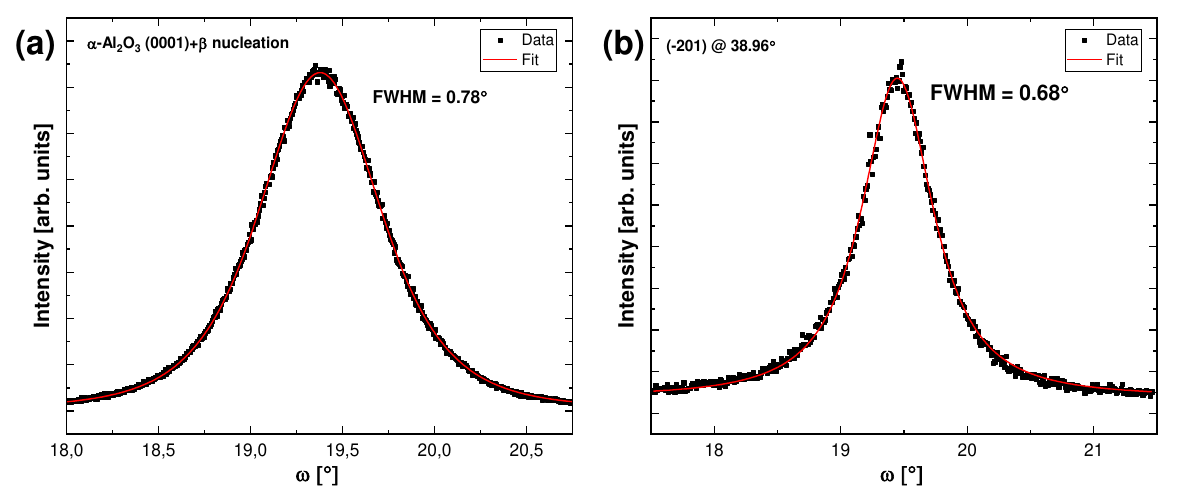}
\caption{\label{Supp5} XRD rocking-curves and respective FWHM of the $\kappa$-Ga$_2$O$_3$ 004 reflex for (a) the sample grown on $\alpha$-Al$_2$O$_3$ with a 70/80 nm of $\beta$-nucleation layer, (b) sample grown on $\beta$-Ga$_2$O$_3$ ($\bar{2}$01)}
\end{figure*}

\begin{figure*}[h!]
\includegraphics[width=0.9\linewidth]{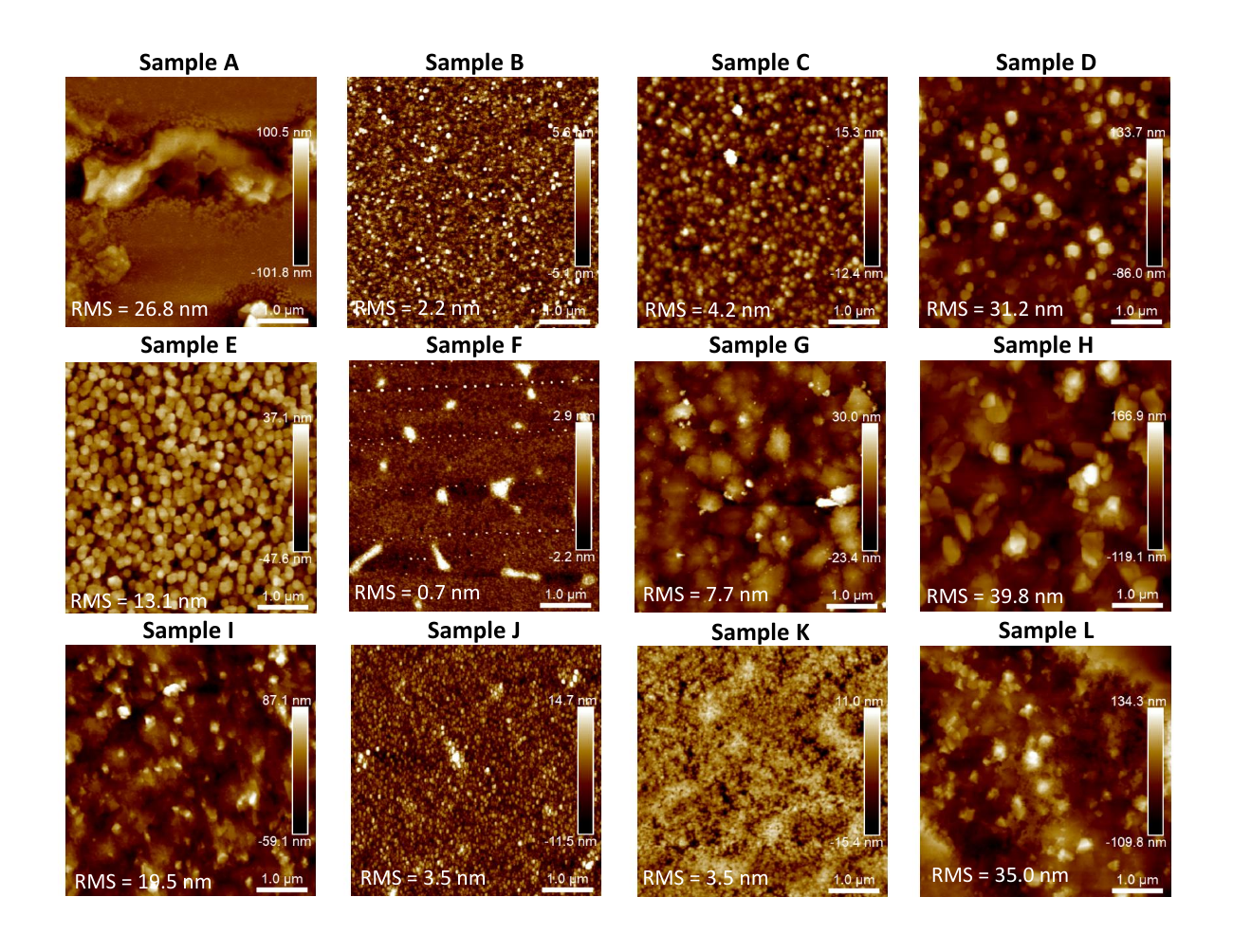}
\caption{\label{Supp6} 5x5 $\mu$m$^2$ AFM images of all the samples grown on the 20/30 nm $\beta$-Ga$_2$O$_3$ ($\bar{2}$01) on $\alpha$-Al$_2$O$_3$ (0001) presented in the work with their relative RMS value.}
\end{figure*}

\begin{figure}

\centering

\rotatebox[origin=c]{90}{\includegraphics[width=1.2\linewidth]{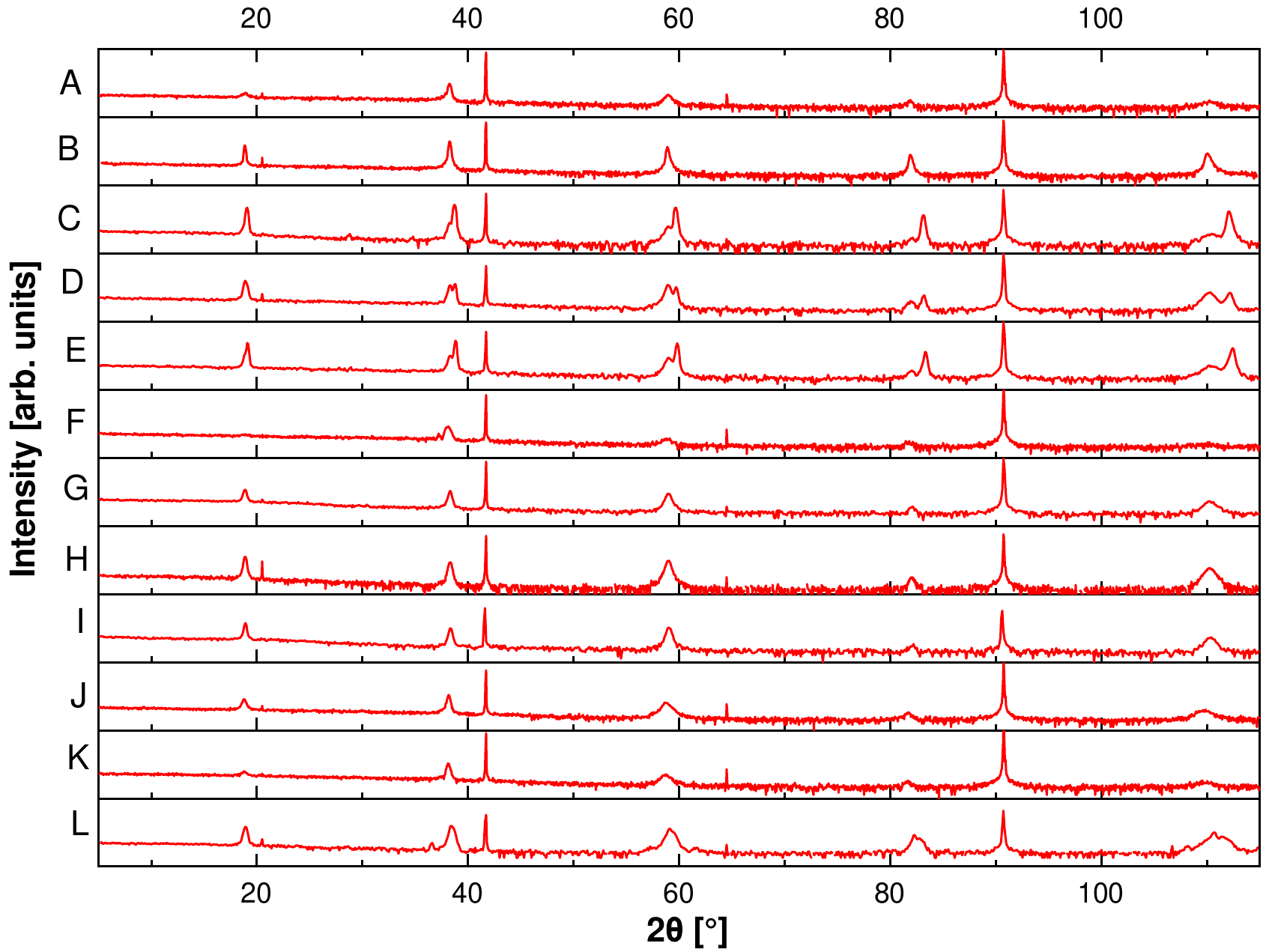}}
\caption{\label{Supp7} XRD 2$\theta$-$\omega$ wide scans for all the samples grown on the 20/30 nm $\beta$-Ga$_2$O$_3$ ($\bar{2}$01) on $\alpha$-Al$_2$O$_3$ (0001) presented in the work.}

\end{figure}

\clearpage
\section*{\centering {Mixed $\beta$/$\kappa$-phase and possible nucleation issues}}

Sample L present a morphology [Fig. S8(a)] which is inhomogeneous on a $\mu$m scale with the presence of large islands with rather flat surfaces and rounded edges rising from a rougher bottom background. 

\begin{figure*}[h!]
\includegraphics[width=0.9\linewidth]{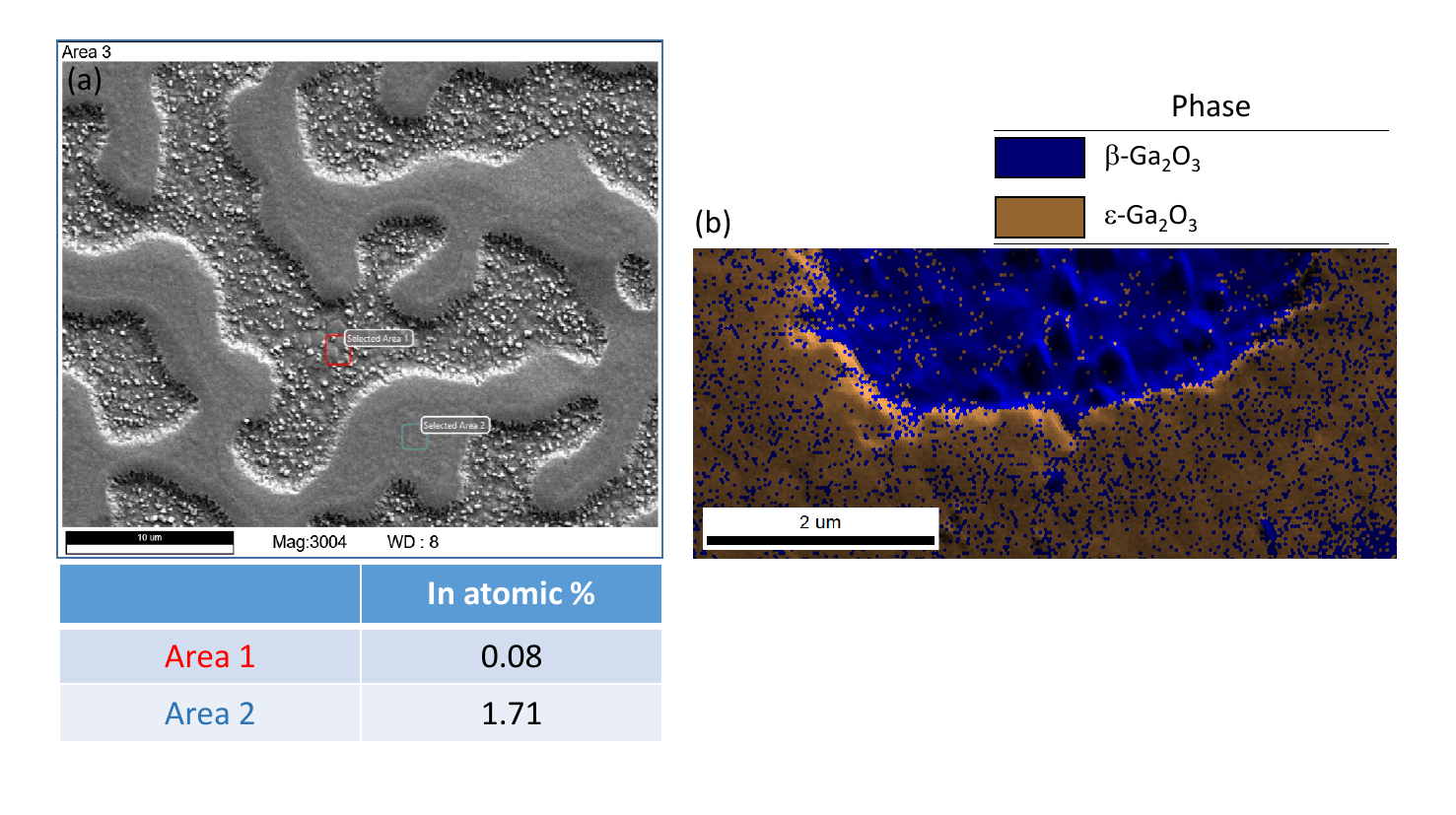}
\caption{\label{Supp8} (a) SEM/EDX image showing the peculiar island formation and the respective In content for the two different morphologies for sample L. (b) EBSD phase map with underlying SEM image image for the $\beta$ (blue) and $\kappa$ (gold) polymorphs.}
\end{figure*}

\begin{figure*}
\includegraphics[width=0.9\linewidth]{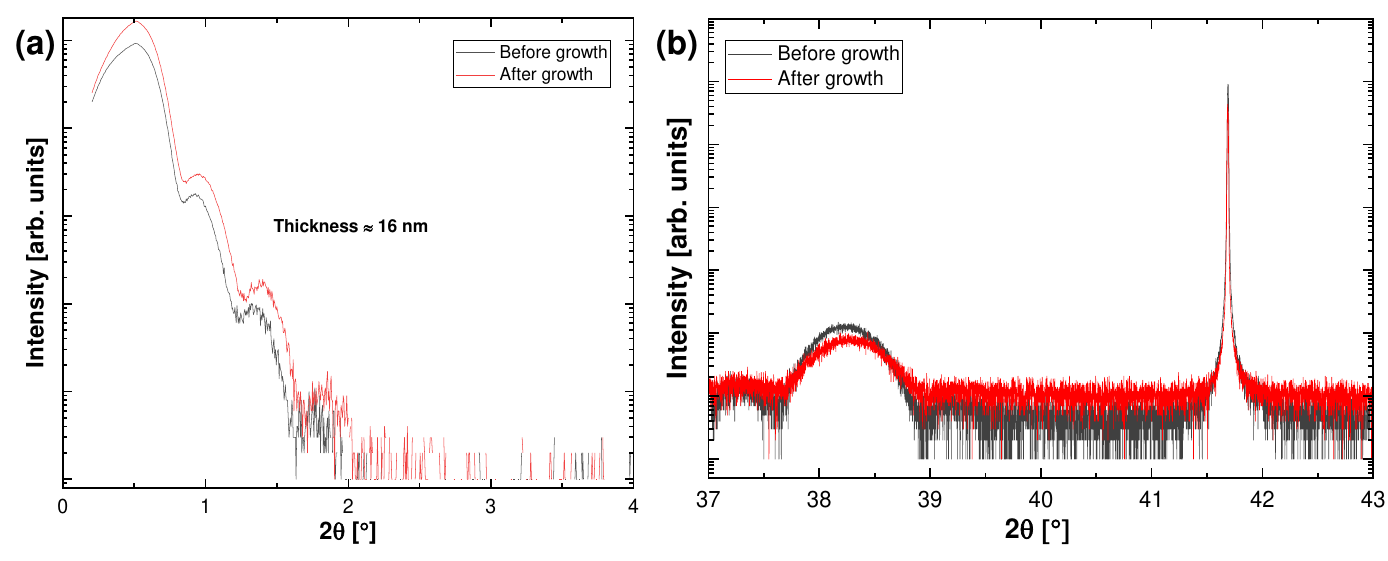}
\caption{\label{Supp9} (a) XRR and (b) 2$\theta$-$\omega$ scan before and after standard PAMBE growth showing that no epilayer has been grown on top of the nucleation layer.}
\end{figure*}

EDX on the two different areas [Fig. \ref{Supp8}(a)] highlight a different In content, showing a higher amount for the flat surface area. Combining SEM and EBSD data [Fig. \ref{Supp8}(b)] shows that the islands are mainly $\kappa$-Ga$_2$O$_3$ (misindexed as $\epsilon$ phase as discussed in the main manuscript), while the rougher background corresponds to $\beta$-Ga$_2$O$_3$. We cannot exclude that in both the depositions in which we achieved a mixed $\beta$/$\kappa$-phase (sample L and D), the deposition conditions should result in pure $\kappa$ stabilization but, due to nucleation issues in the MEXCAT growth, the layer is just growing on a certain area, i.e., the detected $\beta$ fraction is resulting from the $\beta$-Ga$_2$O$_3$ nucleation layer previously deposited without MEXCAT. Alternatively, we can consider the possibility of a competition between the $\beta$- and $\kappa$-phases under such synthesis conditions. In particular, a hypothesis is that growth conditions that allow for the deposition of $\beta$-Ga$_2$O$_3$ without the catalytic element could induce a competition between the “standard” and the In-mediated MEXCAT process, eventually resulting in the mixed $\beta$/$\kappa$ layer. To rule this out, a standard PAMBE growth with the same condition of sample D (but without providing an additional In-flow) was performed on a $\alpha$-Al$_2$O$_3$ (0001) with a 16 nm [Fig. S9(a)] thick $\beta$-Ga$_2$O$_3$  nucleation layer on top. From both in-situ (LR and RHEED) and ex-situ (XRR and XRD) [Fig. \ref{Supp9}(a) and (b)] characterization we did not observe any growth without MEXCAT, therefore excluding the possibility of the two competing growth processes. Another possibility is that initially the $\kappa$-phase is stabilized by the MEXCAT, but the substrate temperature is very close to the one of the phase transition, therefore resulting in a different competition mechanism between the two polymorphs; the $\kappa$ polymorph in fact is thermally stable up to 700 $^{\circ}$C but could show already some minor phase change at 650 $^{\circ}$C; to clarify this point would require additional characterizations which are currently out of the scope of this work.

\end{document}